\documentclass[a4paper,fleqn,usenatbib]{mnras}
\newcommand{\hcn}{L$_{HCN}$}
\newcommand{\mdyn}{M$_{dyn}$}

\usepackage[T1]{fontenc}
\usepackage{ae,aecompl}

\usepackage{epsfig}
\usepackage{amsmath}
\usepackage{graphicx}
\title{Thick Disks, and an Outflow, of Dense Gas in the Nuclei of Nearby Seyfert Galaxies}\author[M.-Y. Lin et al.]{Ming-Yi~Lin$^1$\thanks{E-mail: acdo2002@gmail.com},
R.I.~Davies$^1$,
L.~Burtscher$^1$,
A.~Contursi$^1$,
R.~Genzel$^1$, 
\newauthor E.~Gonz\'alez-Alfonso$^2$,
J.~Graci\'a-Carpio$^1$,
A.~Janssen$^1$,
D.~Lutz$^1$,
G.~Orban~de~Xivry$^1$,
\newauthor D.~Rosario$^1$,
A.~Schnorr-M\"uller$^1$,
A.~Sternberg$^3$,
E.~Sturm$^1$,
L.~Tacconi$^1$\\
$^1$Max-Planck-Institut f\"ur extraterrestrische Physik, Postfach 1312, 85741, Garching, Germany\\
$^2$Universidad de Alcal\'a, Departamento de F\'isica y Matem\'aticas, Campus Universitario, E-28871 Alcal\'a de Henares, Madrid, Spain\\
$^3$Raymond and Beverly Sackler School of Physics \& Astronomy, Tel Aviv University, Ramat Aviv 69978, Israel}

\date{Accepted 2016 February 18. Received 2016 February 18; in original form 2015 September 24}

\pubyear{2015}

\begin{document}
\label{firstpage}
\pagerange{\pageref{firstpage}--\pageref{lastpage}}
\maketitle

\begin{abstract}
We discuss the dense molecular gas in central regions of nearby Seyfert galaxies, and report new arcsec resolution observations of HCN\,(1-0) and HCO$^+$\,(1-0) for 3 objects.
In NGC\,3079 the lines show complex profiles as a result of self-absorption and saturated continuum absorption.
H$^{13}$CN reveals the continuum absorption profile, with a peak close to the galaxy's systemic velocity that traces disk rotation, and a second feature with a blue wing extending to $-350$\,km\,s$^{-1}$ that most likely traces a nuclear outflow.
The morphological and spectral properties of the emission lines allow us to constrain the dense gas dynamics.
We combine our kinematic analysis for these 3 objects, as well as another with archival data, with a previous comparable analysis of 4 other objects, to create a sample of 8 Seyferts.
In 7 of these, the emission line kinematics imply thick disk structures on radial scales of $\sim$100\,pc, suggesting such structures are a common occurrence.
We find a relation between the circumnuclear \hcn\ and \mdyn\ that can be explained by a gas fraction of 10\% and a conversion factor $\alpha_{HCN} \sim 10$ between gas mass and HCN luminosity. Finally, adopting a different perspective to probe the physical properties of the gas around AGN, we report on an analysis of molecular line ratios which indicates that the clouds in this region are not self-gravitating.
\end{abstract}

\begin{keywords}
Galaxies: nuclei -- 
galaxies: active -- 
galaxies: ISM --
galaxies: Seyfert --
submillimetre: galaxies
\end{keywords}

\section{Introduction}
\label{sec:intro}

The standard unification scheme for active galactic nuclei (AGN) proposes that the intrinsic properties of AGN are similar, and the disparity of observational properties arises from the different viewing angles with respect to an obscuring medium, resulting in type 1 and type 2 AGNs \citep{Antonucci1993, UP1995}.
The reality may be more complex than this simple picture suggests \citep{net15} although there is general agreement that the obscuring medium of gas and dust is a geometrically and optically thick toroidal structure (perhaps including an outflowing wind, e.g. \citealt{Elitzur2006}) surrounding the AGN accretion disk, with its inner edge at the dust sublimation radius.
Numerous observations confirm that there is dust at temperatures in the range 1000--1500\,K (e.g. see \citealt{bur15} and references therein).
And a variety of models for smooth \citep{pie92,Granato1994,Granato1997,sch05,fri06} and clumpy \citep{Krolik1988,Nenkova2002,hoe06,sch08,sch14} media have been constructed to reproduce the spectral energy distribution (SED) at infrared wavelengths as well as the silicate feature at $\sim10$\micron\ \citep{Schweitzer2008}.
However, a number of difficulties remain when applying these models to observations.
\cite{Feltre2012} point out that in addition to the parameters describing the physical geometry, the assumptions built into the models can have a major impact on the resulting SED.
And in the best studied cases where interferometry can spatially resolve some of the structure, a component that is aligned along a direction consistent with an outflow appears to be responsible for much of the mid-infrared continuum \citep{Tristram2014}.
Finally, in the context of the observations we present here, the outer edge of the torus is poorly defined for the majority of these models, because beyond a few tens of parsecs it contributes relatively little to the warm dust continuum against which the models are tested.

On radial scales of tens to a hundred parsecs, it is not clear whether one should still refer to the gas and dust structure as a torus or a (circum-)nuclear molecular disk.
There have been claims that even on these scales the molecular disk is thick enough to have an impact on obscuration towards the AGN and so contributes to the observed difference between type~1 and type~2 AGN \citep{Hicks2009,Sani2012}; and hydrodynamical simulations suggest that turbulence injected via supernova from recent or on-going star formation can puff up the disk to heights of $\sim10$\,pc \citep{wad09,wad12}.
Spatially resolving molecular tracers on these scales in nearby AGN has become possible during the last decade, through the use of adaptive optics systems operating at near-infrared wavelengths and interferometers working in the millimetre regime.

\citet{Hicks2009} observed the $\sim2000$\,K H$_{2}$ via the 2.12 $\micron$ 1-0\,S(1) line.
In most cases the velocity dispersion ($\sigma_{H2}$) of this line exceeded its rotational velocity ($v$) on radial scales of $\sim 50$\,pc.
Their conclusion was that the disk is relatively thick.
The 2.12 $\micron$ line traces a small excited fraction of the H$_{2}$ mass that might have peculiar kinematics, so that kinematics from a tracer of the bulk of the molecular gas is desirable.
Using the 3\,mm HCN\,(1-0) line, \citet{Sani2012} modeled the emission as a rotating disk, and also found that a large intrinsic velocity dispersion ($\sigma_{HCN}$) was required to match the observations.
Intriguingly, these results together, $\sigma_{H2} \sim 50$--100\,km\,s$^{-1}$ and $\sigma_{HCN} \sim 20$--40\,km\,s$^{-1}$, suggest that the molecular gas is stratified with the warmest gas being the most turbulent and reaching to larger scale heights while the denser clouds remain closer to the disk plane.
Although several observations (e.g. \citealt{Hicks2009} and \citealt{Muller2013}) have confirmed the presence of thick disks with $\sigma/v \sim 0.4$ on radial scales of $\sim50$\,pc in some objects, it is not clear how the kinetic energy is continuously supplied in order to maintain the vertical structure of a rotating disk.
One plausible explanation is that nuclear star formation can provide the necessary energy.
Stellar radiation pressure \citep{Thompson2005} is one option if there are sufficient OB stars and the ISM is optically thick.
However, while some observations are consistent with on-going star formation \citep{esq14,rif09}, others point more towards post-starburst populations \citep{cid04,Davies2007}, and in some cases there is clear evidence that there cannot be on-going star formation in the central tens of parsecs \citep{hic13}.
Alternatively, supernova explosions, associated even with a very modest star formation rate, are able to generate sufficient turbulence in the disk to yield $\sigma \sim 20$\,km\,s$^{-1}$ to radii of 25\,pc \citep{Wada2002,wad09}.
A third possibility is that the gas flows induced by disk instabilities toward inner radii can also maintain a thick disk \citep{Hopkins2012}.
\cite{vol08} also argued that the energy from inflowing gas could in principle be used to thicken the central disk.

The main purpose of this work is to model the intrinsic geometry (i.e. radial and height scales) of the molecular disk in the central $\sim100$\,pc around AGN, by using high resolution HCN\,(1-0) and HCO$^+$\,(1-0) observations.
In addition, we use Large Velocity Gradient (LVG) calculations for molecular line ratios to constrain the cloud properties in one of the objects, NGC\,6951.

The HCN and HCO$^+$ molecules are ideal indicators of dense gas, since the higher dipole moments of their 1-0 rotational transitions can trace $\sim 100$--500 times denser gas than the same rotational transitions of the CO molecule, and are sensitive to gas at  $n_{H2} \sim 10^{4-5}$\,cm$^{-3}$ \citep{Papadopoulos2007}. 
These molecules have been successfully observed not only in dense star-forming regions but also in a number of Seyfert galaxies \citep{Krips2008,koh05,Kohno2008,Sani2012}, although typically at resolutions of 5--20\,arcsec.
In the nuclear region of Seyfert galaxies, HCN\,(1-0) and HCO$^+$\,(1-0) can be excited either by UV light from the accretion disk or X-ray radiation from the corona \citep[and the references therein]{Sternberg1995,mal96,Lepp1996,Meijerink2005,bog05,Meijerink2007}.

This paper is organized as follows.
The HCN\,(1-0) and HCO$^+$\,(1-0) observations of NGC\,3079, NGC\,6764, and NGC\,5033 are described in Section~\ref{sec:Obs}. 
We present the observed molecular gas distribution and kinematics for each galaxy in Section~\ref{sec:allR}.
We apply a simple dynamical model to the emission lines in Section~\ref{sec:model_kinematics}, and in section~\ref{sec:cd3079} use the H$^{13}$CN\,(1-0) absorption to estimate the column density of the disk in NGC\,3079.
Section~\ref{sec:hcn_mdyn} brings in data from the literature to explore the relation between HCN luminosity and dynamical mass for a sample of 8 objects with high spatial resolution HCN observations.
Finally, Section~\ref{sec:ngc6951} looks more closely at the cloud properties for one specific object NGC\,6951 where data for suitable transitions are available.
We summarize our conclusions in Section~\ref{sec:conc}.

\begin{table*}
\begin{center}
  \caption{Summary of observations for the 3 Seyfert galaxies. \newline
(1) Source name; (2) Beam size; (3) Beam position angle; (4) Channel width; (5) noise per beam; (6) Distance; (7) Physical scale of 1\arcsec.}

    \begin{tabular}{*{7}{c}} \hline    
      Source & Beam size  &  P.A.  & Channel resolution   &  Noise & Distance  & scale \\
      &  &    (degree) &  (km s$^{-1}$) & (mJy\,beam$^{-1}$\,ch$^{-1}$) & (Mpc)  & (pc/\arcsec) \\ 
  \hline \hline           
 NGC 3079 & 1.17$\arcsec$ $\times$ 0.94$\arcsec$ & 167 & 17.1 & 0.37 & 20 & 85 \\
 NGC 6764 & 1.16$\arcsec$ $\times$ 0.78$\arcsec$ &  13 & 34.4 & 0.29 & 32 & 150 \\ 
 NGC 5033 & 1.13$\arcsec$ $\times$ 0.83$\arcsec$ &  38 & 34.2 & 0.31 & 19 & 73 \\
  \hline
        \end{tabular}
 \\
%           *************Noise: root mean square of uncertainty for each observation. is dependent on calibration??? (binning should be consider???).
  %         ?0.65? is for 10MHz spectral resolution, but we use 5MHz $->$ ?0.65/2???
  
%$^{(a)}$Effective area covered by the survey; $^{(b)}$ Flux limit of the catalog; $^{(c)}$Total number of sources; $^{(d)}$Number of sources cross-matched with the $70\micron$ catalog;  $^{(e)}$References describing the catalog; $^{(f)}$Catalog concatenated from XMM- and Chandra-catalogs; sources matched between the two catalogs are not duplicated; the cross-match between this catalog and the $70\micron$ catalog constitute our primary  catalog.
\label{tab:obs_info}   
\end{center}
\end{table*}

\section{Observations}
\label{sec:Obs}

We used the six 15-m antennas of the IRAM Plateau de Bure interferometer (PdBI) to observe NGC\,3079, NGC\,6764, and NGC\,5033 at 88\,GHz (3\,mm) with the WideX correlator in the extended A configuration. 
The basic calibration steps of three galaxies were done with the CLIC software. 
The observations for the individual galaxies are described below and summarised in Table~\ref{tab:obs_info}.

Observations of NGC\,3079 (systemic velocity 1147\,km\,s$^{-1}$ and inclination 77$\degr$, \citealt{Koda2002}) were carried out on 19\,January\,2011 for programme UD8A.
The receiver band was centered at 87.55\,GHz with its 3.6\,GHz bandwidth covered all features from HCO$^+$\,(1-0) at 89.2\,GHz to H$^{13}$CN\,(1-0) at 86.3\,GHz (rest frequencies).
During the 5.5\,hr track the average antenna efficiency was 22.6\,Jy/K, and the precipitable water vapour (pwv) improved during the track from 2--4\,mm to $<2$\,mm.
The calibrators include 3C273, 0923+392, 0954+556, and MWC349. 
After data reduction, the synthesised beam at 89\,GHz was $1.2\arcsec \times 0.9\arcsec$ at position angle (PA) $167\deg$. 
The data were binned to a spectral resolution of 5\,MHz corresponding to a channel width of 17\,km\,s$^{-1}$, yielding a root mean square uncertainty of 0.37\,mJy/beam.
This was chosen in order to balance signal-to-noise with high spectral resolution since some of the absorption features are much narrower than the emission lines.
Images of the channels were reconstructed with a pixel size of 0.2\arcsec.
We adopt a distance to NGC\,3079 of 19.7\,Mpc, for which 1$\arcsec$ corresponds to 85\,pc.

NGC\,6764 was also observed on 19\,January\,2011, after NGC\,3079, with the receiver band centered at 87.15\,GHz.
This setting was based on the systemic velocity of 2416\,km\,s$^{-1}$ used by \cite{hot06}, who found deep H\,I absorption on the location of the nucleus at a velocity of 2426\,km\,s$^{-1}$ consistent with that velocity.
However, the profiles of both the HCN\,(1-0) and HCO$^+$\,(1-0) lines in our data are centered at 2468\,km\,s$^{-1}$ (each differing by only 10\,km\,s$^{-1}$ from that mean).
This implies that the H\,I may be outflowing, a phenomenon that has been observed in a number of galaxies (see \citealt{mor12} for a review), for example in the Seyfert IC\,5063 the H\,I absorption is blue-shifted with respect to the H\,I emission with velocities comparable to a blue wing in the CO\,(2-1) line \citep{mor07,mor15}. 
In our analysis we use 2468\,km\,s$^{-1}$ as the systemic velocity for NGC\,6764.
We also adopt an inclination of 62$\degr$ based on CO\,(1-0) kinematics \citep{Leon2007}.
During the 4.9\,hr track the average antenna efficiency was 22.9\,Jy/K, and the pwv was $<1$\,mm. 
The calibrators include MWC349, 1954+513, 1739+522, 1749+096, and 1823+568.
The synthesised beam at 89\,GHz in the processed data was $1.2\arcsec \times 0.8\arcsec$ at a PA of $13\deg$.
The data were spectrally binned to 10\,MHz, corresponding to a channel width of 34\,km\,s$^{-1}$ and a root mean square uncertainty of 0.29\,mJy/beam.
The channel binning is coarser than for NGC\,3079 but sufficient to sample the velocity width of the emission lines.
In order to fit disk models, the data were converted to an image plane with a pixel size of 0.3\arcsec.
We adopt a distance to NGC\,6764 of 31.7\,Mpc, for which 1$\arcsec$ corresponds to 150\,pc.  

Observations of NGC\,5033 (systemic velocity 875\,km\,s$^{-1}$, \citealt{huc95}; inclination 68$\degr$, \citealt{Thean1997} and \citealt{Kohno2003}) were conducted over two nights on 26 and 28\,January\,2011, with tracks of 3.1\,hrs and 3.9\,hrs respectively.
The receiver band was centered at 87.60\,GHz.
The average antenna efficiency of 22.7\,Jy/K, with pwv $<1$\,mm on the first night and $<3$\,mm on the second night except near the end when it increased slightly.
The calibrators include 3C273, MWC349, 1308+326, and 0355+508.
After data reduction, the synthesised beam size at 89\,GHz was $1.1\arcsec \times 0.8\arcsec$ at a PA of $38\deg$. 
The reconstructed data for NGC\,5033 share the same spectral binning and pixel size as NGC\,6764, and have a root mean square uncertainty of 0.31\,mJy/beam.
We adopt a distance to NGC\,5033 of 18.7\,Mpc, for which 1$\arcsec$ corresponds to 73\,pc. 

We also make use of HCN\,(1-0) data for NGC\,7469 presented in \cite{Davies2004}, which have a beam size of $2.0\arcsec$. 
We adopt a distance of 58\,Mpc and an inclination of 45$\degr$.

\begin{table*}
%\begin{center}
%\hspace{-5.0cm}
 \caption{Summary of the continuum properties for the 3 targets: \newline 
(1) Source name; (2) Flux density; (3) The observed FWHM of the major and minor axes; (4) Position angle (east of north).}
    \begin{tabular}{cccc} \hline               
     Source & Flux density  &  Major $\times$ Minor axis & P.A.  \\
      & (mJy) & (arcsec$^2$) &  ($\degr$)  \\ 
  \hline \hline
  NGC 3079  & 27.8 $\pm$ 0.15  &  1.19 $\pm$ 0.01 $\times$ 1.06 $\pm$ 0.01 & 176 $\pm$ 1   \\
   NGC 6764  & 0.6 $\pm$ 0.06  &  1.04 $\pm$ 0.05 $\times$ 0.82 $\pm$ 0.04 & 8 $\pm$ 2   \\
  NGC 5033  & 0.7 $\pm$ 0.08  &  1.01 $\pm$ 0.04 $\times$ 0.79 $\pm$ 0.03 & 41 $\pm$ 2    \\\hline
    \end{tabular} 
    \label{tab:conti_info}    
%    \end{center}
\end{table*}

\begin{table*}
%\begin{center}
  \caption{The observed properties of molecular emission for 3 targets: \newline
(1) Source name; (2) Molecule; (3) Flux$^{a}$; (4) Observed FWHM of major and minor axes; (5) Position angle (east of north); (6) Separation between centers of red and blue channel maps; (7) Position angle between centers of the red and blue channel maps; (8) Spectral FWHM of line (3\arcsec\ aperture).}
    \begin{tabular}{*{8}{c}} \hline               
     Source & Line & Flux   &  Major $\times$ minor axis & PA & b/r sep. & PA$_{b/r}$ & Line width  \\
      & & (Jy km s$^{-1}$) & (arcsec$^{2}$) &  ($\degr$) & (arcsec) & ($\degr$) & (km s$^{-1}$) \\ 
  \hline \hline
  NGC 3079  &HCN &  8.20$^{b}$  &  (2.94 $\pm$ 0.14) $\times$ (1.44 $\pm$ 0.06)  & 167 $\pm$ 2  & 1.51 $\pm$ 0.02 & -21 $\pm$ 2 & 360$^{c}$  \\  
                    &HCO+& 4.78$^{b}$  &   strong absorption  &  strong absorption & 1.70 $\pm$ 0.04 & -20 $\pm$ 2  & - \\ \hline
 NGC 7469  &HCN & 5.30 $\pm$ 0.10 &  (3.43 $\pm$ 0.10) $\times$ (2.73 $\pm$ 0.08)  & 78 $\pm$ 15 & 1.14 $\pm$ 0.04 & 118 $\pm$ 2 & 236 $\pm$ 9 \\
 NGC 6764 & HCN & 1.92 $\pm$ 0.11  &  (1.71 $\pm$ 0.14) $\times$ (1.31 $\pm$ 0.09)  & -54 $\pm$ 25 & 0.82 $\pm$ 0.11 & -87 $\pm$ 7 & 214 $\pm$ 17 \\
                         &HCO+ & 2.28 $\pm$ 0.09  &  (1.69 $\pm$ 0.11) $\times$ (1.21 $\pm$ 0.06)  & -66 $\pm$ 7 & 1.04 $\pm$ 0.11 & -85 $\pm$ 8 & 203 $\pm$ 16 \\
 NGC 5033  & HCN & 1.16 $\pm$ 0.14  &  (1.97 $\pm$ 0.30) $\times$ (1.09 $\pm$ 0.13) & -17 $\pm$ 10 & 1.06 $\pm$ 0.15 & 158 $\pm$ 19 & 181 $\pm$ 32  \\
                     &HCO+ & 0.71 $\pm$ 0.14  &  (1.76 $\pm$ 0.40) $\times$ (1.22 $\pm$ 0.20) & 52 $\pm$ 67  & 0.99 $\pm$ 0.56 & 184 $\pm$ 51 & 185 $\pm$ 71 \\\hline
    \end{tabular} \\
    \label{tab:emi_info}    
$^{a}$ These are given in a 3\arcsec\ aperture, except for NGC\,3079 where we have used a 5\arcsec\ aperture.\\
$^{b}$ Fluxes are integrated across the observed line profile above the continuum level. Applying corrections in a simple way as illustrated by the blue line in the right panel of Figure~\ref{fig:3079profiles} yields fluxes for HCN\,(1-0) of 16.1\,Jy\,km\,s$^{-1}$ when accounting for only the continuum absorption, and 18.7\,Jy\,km\,s$^{-1}$ when also accounting for the self-absorption. The equivalent corrected fluxes for the HCO$^+$\,(1-0) line are 13.3 and 19.5\,Jy\,km\,s$^{-1}$.\\
$^{c}$ Estimated intrinsic line width after correcting for continuum absorption and self-absorption (i.e. it corresponds to the blue Gaussian line profile in the right panel of Fig~\ref{fig:3079profiles}). We have adopted an uncertainty of 20\,km\,s$^{-1}$ typical of the other linewidth measurements.
\end{table*}

\section{Gas distribution and kinematics}
\label{sec:allR}

We begin this section by describing the general properties of the three galaxies, and then present the details for each individual source in the following subsections. 
HCN\,(1-0) at 88.63\,GHz, and HCO$^+$\,(1-0) at 89.19\,GHz were detected in all galaxies.

In every case, the 3\,mm continuum (spectrally integrated over channels that are free from molecular transitions) is seen as a single compact source.
Since the major and minor axes, and the position angle, are very close to those of the beam, the continuum appears to be spatially unresolved.
The flux density and morphology of the continuum sources are derived from line-free regions of the spectrum, and are given in Table~\ref{tab:conti_info}. 

In contrast, the HCN\,(1-0) and HCO$^+$\,(1-0) emission line morphologies are spatially extended, although by not more than a few arcsec.
We define the centre from the continuum image, and extract an integrated spectrum in a 3\arcsec\ diameter aperture around that.
The flux density of the emission lines is derived from this by summing over all channels across the spectral line profile.
It is important to note that for NGC\,3079, HCN\,(1-0) and HCO$^+$\,(1-0) both show a clear P-Cygni shape, with absorption cutting through the emission line profile.
For both NGC\,6764 and NGC\,5033, the lines are seen in emission only.
As for the continuum, the morphology of the molecular lines is quantified by fitting a two-dimensional Gaussian function to the line map.
In addition, we have measured the centers of the emission summed over red channels and blue channels separately (the velocity ranges are from line centre to $\pm350$--480\,km\,s$^{-1}$ depending on the line and object; specific ranges are given in the following subsections), in order to derive their relative separation and position angle.
The observed properties of the molecular emission lines are summarized in Table~\ref{tab:emi_info}.

\begin{figure*}
\begin{center}
  %\hspace{-1.15cm}
%  \vspace{-1.0cm}
      \includegraphics[width=120mm]{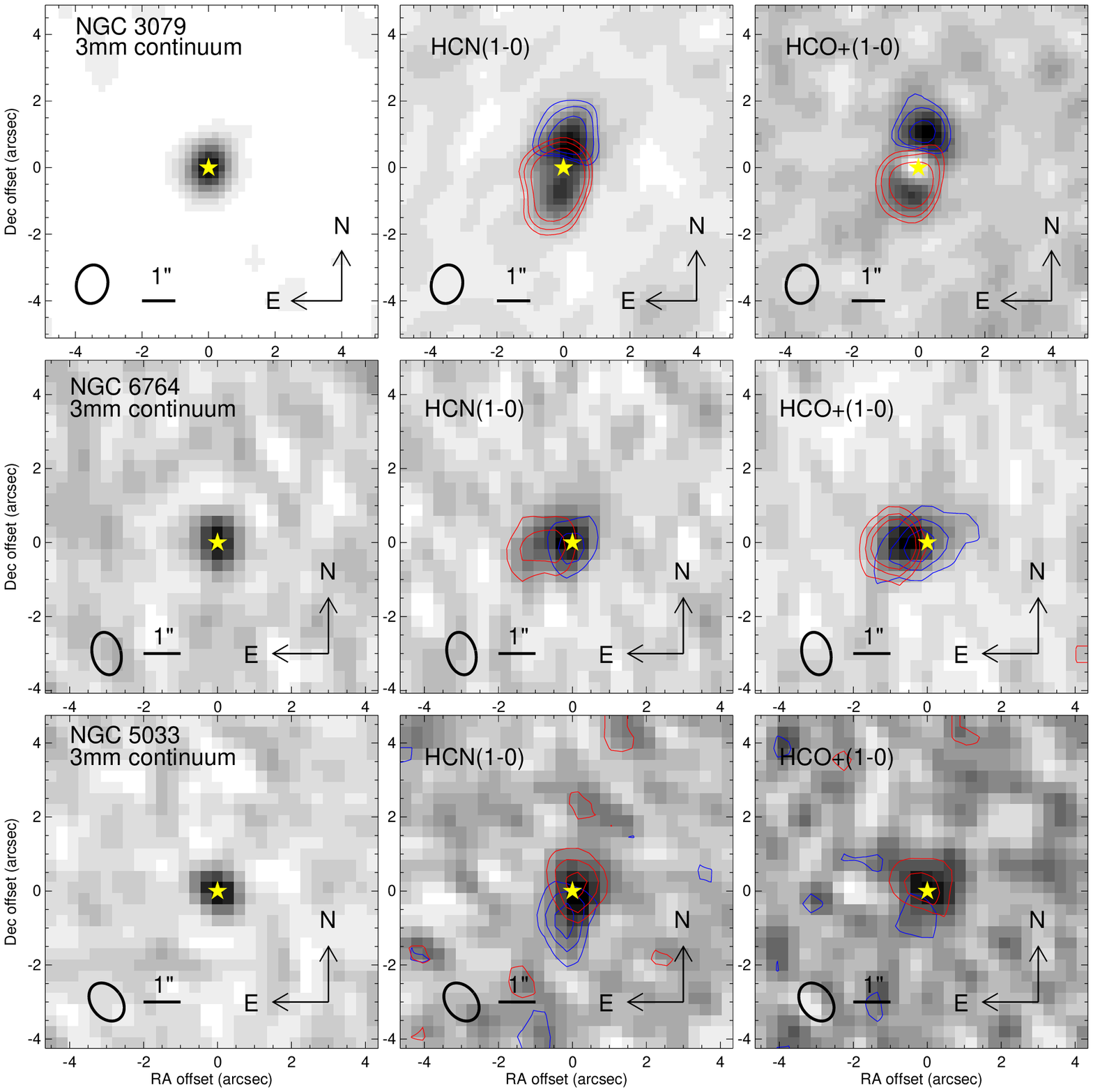}
         \caption{The 3\,mm continuum (left panels), integrated HCN\,(1-0) (middle panels), and integrated HCO$^+$\,(1-0) (right panels) emission maps of NGC\,3079 (top row), NGC\,6764 (middle row), and NGC\,5033 (bottom row). 
In each panel, the peak position of 3\,mm continuum is marked as a yellow star and the synthesized beam is drawn as an ellipse. 
The 3\,mm continuum has been subtracted from HCN and HCO$^+$ maps.
The blue and red contours represent the distribution of integrated blue and red channels of the emission lines, and are easily spatially resolved.
For NGC\,3079, the map is centered at 10\,01\,57.80~$+$55\,40\,47.2 (J2000).
The absorption appears as a `hole' in the central region, more clearly in the HCO$^+$ map due to the stronger absorption of that line.
Blue and red contours are drawn at levels of 2$\sigma$, 3$\sigma$, and 5$\sigma$.
Note that while the absorption is blue-shifted, in these images it appears to be occuring within the red contours. As becomes clear in Figure~\ref{fig:PVdiagram}, this is because the blue-shifted emission is filling in the absorption and therefore remains below the continuum level, so the blue contours do not extend over that region. 
For NGC\,6764, the map is centered at 19\,08\,16.38~$+$50\,55\,59.4 (J2000), and blue and red contours are drawn at levels of 0.5$\sigma$, 1$\sigma$, and 1.5$\sigma$.
For NGC\,5033, the map is centered at 13\,13\,27.47~$+$36\,35\,37.9 (J2000), and blue and red contours are drawn at levels of 0.5$\sigma$, 1$\sigma$, and 1.5$\sigma$.}
\label{fig:conhcnhco}
\end{center} 
\end{figure*}

\begin{figure*}
\begin{center}
  \includegraphics[width=175mm]{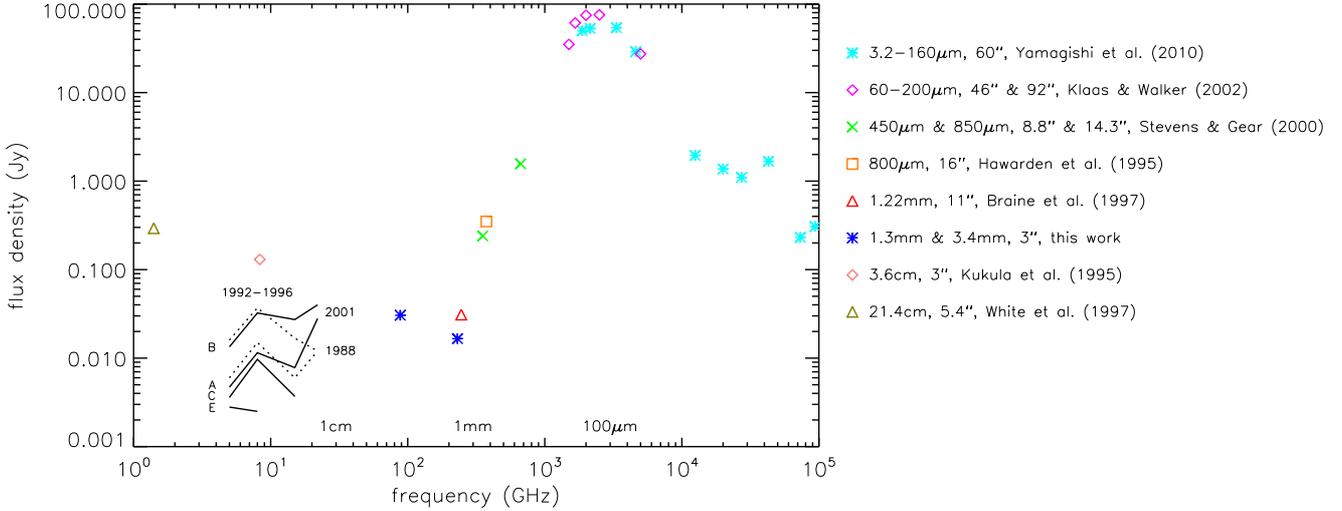}
  \caption{Radio to infrared spectral energy distribution (SED) for NGC\,3079 as well as the interferometric mm-to-cm SED for the 4 radio emitting knots.
Knots A and C are associated with approaching jet to the Southeast (SE) that is in front of the nuclear disk; while knot B is associated with the receeding jet to the Northwest (NW) that is behind the disk.
The SED shows that the continuum longward of 1\,mm is non-thermal, and that the 3\,mm continuum in the central few arcsec is likely dominated by the two radio knots A and B.
References for photometry: 
3.2--160\,$\mu$m (60\arcsec\ apertures) from \citet{yam10};
60--200\,$\mu$m (46\arcsec\ and 92\arcsec apertures) from \citet{kla02};
450\,$\mu$m (8.8\arcsec) and 850\,$\mu$m (14.3\arcsec) from \citet{ste00};
800\,$\mu$m (16\arcsec) from \citet{haw95};
1.22\,mm (11\arcsec) from \citet{bra97};
1.3\,mm and 3.4\,mm (3\arcsec\ aperture) from this work;
3.6\,cm (3\arcsec\ beam) from \citet{kuk95};
21.4\,cm (5.4\arcsec\ beam) from \citet{whi97}.
The interferometric radio continuum measurements of components A, B, C, and E with beam sizes of 0.3--8\,mas are from \citet{Trotter1998} and \citet{Kondratko2005}.
}
\label{fig:cont3079}
\end{center} 
\end{figure*}

\begin{figure*}
\begin{center}
  %\hspace{-0.3cm}
  %\vspace{-0.75cm}
 \includegraphics[width=160mm]{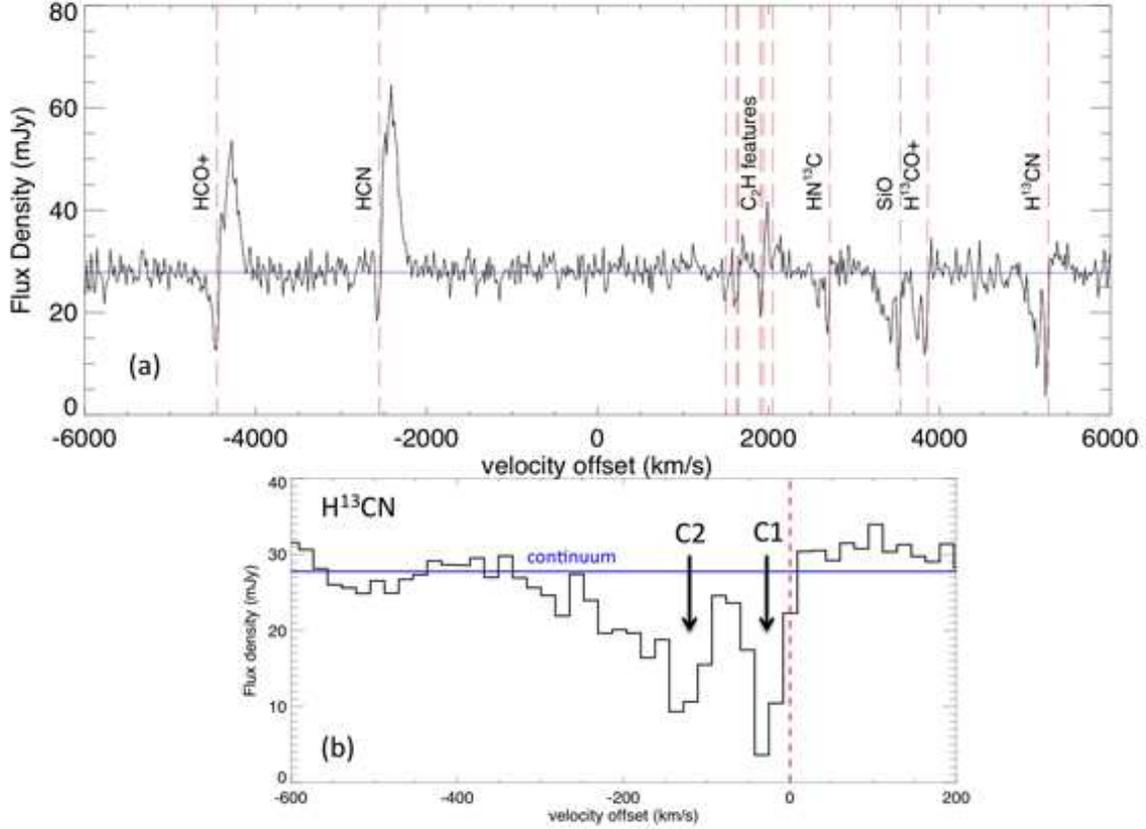}
 \caption{Panel (a): the spectrum of NGC\,3079 integrated within a $3\arcsec$ aperture (velocity offset is with respect to the band centre). 
Transitions of H$^{12}$CO$^+$\,(1-0), H$^{12}$CN\,(1-0), HN$^{13}$C\,(1-0), SiO\,(2-1), H$^{13}$CO$^+$\,(1-0), and H$^{13}$CN\,(1-0) are detected, as well as hyperfine transitions of C$_{2}$H\,(1-0). 
The blue solid line is a constant representing the 3\,mm continuum. 
The red dash lines indicate the velocity offsets of each molecular transition with respect to the galaxy systematic velocity of 1147\,km\,s$^{-1}$.
Panel (b): The complex absorption of the H$^{13}$CN\,(1-0) transition (in a 3\arcsec\ aperture). 
Velocity offsets are given with respect to its systemic, denoted by the dashed red line.
The deepest absorption is labeled as ``C1'' and the broader absorption is labeled as ``C2''.
Note that both absorption features are blue-shifted.}
\label{fig:ngc3079spec}
\end{center}  
\end{figure*}

\begin{figure*}
\begin{center}
      \includegraphics[width=120mm]{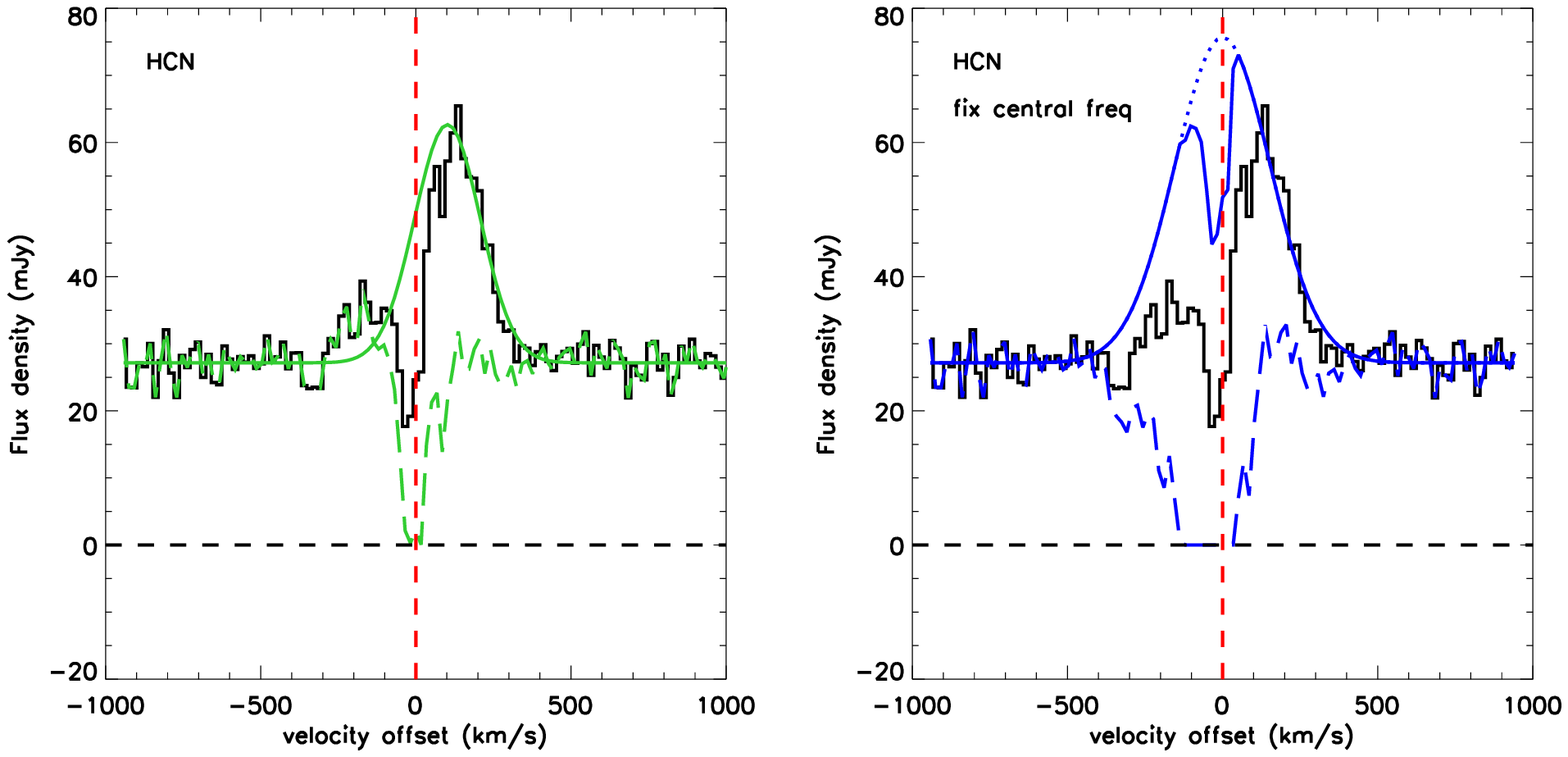}      
         \caption{Observed and reconstructed HCN\,(1-0) line profiles in NGC\,3079.
In both panels the observed profile is drawn as a black line, and the reconstructions assume that this is the sum of continuum absorption (occuring over spatial scales of a few milliarcsec associated with radio knot B) and a Gaussian emission line (originating in the disk on scales of several arcsec).
Left: the intrinsic profile has been reconstructed with the requirements that the red side of the observed profile is unabsorbed, and that the continuum absorption (dashed green line) cannot go below zero. The resulting intrinsic emission line (solid green line) is red-shifted with respect to systemic and also under-estimates the weak emission observed on the blue side of the line.
In addition the continuum absorption lacks the blue wing seen in the H$^{13}$CN profile.
Right: the additional requirement that the intrinsic line profile must be centered at the systemic velocity has been applied. The continuum absorption (dashed blue line) is now much broader and extends to velocities of -400\,km\,s consistent with the absortption profile of the H$^{13}$CN\,(1-0) line. However, to match the observed line profile, the intrinsic Gaussian emission line (dotted blue line) must be self-absorbed (solid blue line). This is plausible because the disk of NGC\,3079 is close to edge-on, so there is considerable gas at large scales in the disk along the line of sight to the nucleus. In addition, the self-absorption is centered roughly at systemic, which would be expected if cool gas within the disk is causing the self-absorption.}
\label{fig:3079profiles}
\end{center} 
\end{figure*}

\subsection{NGC\,3079}
\label{sec:Rngc3079}

The 3\,mm continuum is displayed in the left panel of Figure~\ref{fig:conhcnhco}.
A Gaussian fit indicates FWHM of $1.19\arcsec \times 1.04\arcsec$ at a PA of 176$\degr$.
These are comparable to the beam size, indicating that the source is not spatially resolved, and has an intrinsic size $<0.5$\arcsec.
Because of the absorption features we have observed, the origin of the 3\,mm continuum is an important issue.
Maps of the radio continuum at 5--22\,GHz and at resolutions to $<1$\,mas \citep{Trotter1998,Kondratko2005,Middelberg2007} demonstrate that it is dominated by regions that are 1--3\,pc from the dynamical centre as defined by \cite{Trotter1998}.
Figure~\ref{fig:cont3079} shows the radio to infrared SED for NGC\,3079 and indicates that the two radio lobes A and B are likely to dominate also the 3\,mm continuum.
In 2001, the sum of the flux densities of these two components (68\,mJy) at 22\,GHz was equal to the total radio continuum integrated over arcsec scales; 
and there is some evidence that they have continued to brighten at 22\,GHz in subsequent years \citep{Middelberg2007}.
The discussion below is based on the conclusion that the 3\,mm continuum, which is clearly non-thermal, is dominated by the radio components A and B, rather than by any emission from the dynamical centre.

The maps of HCN\,(1-0) and HCO$^+$\,(1-0) are presented in the middle and right panels of Figure~\ref{fig:conhcnhco}.
The grey scale image indicates the full line distribution, while the red and blue contours represent the distribution of integrated red channels and blue channels of the emission lines.
For NGC\,3079, the ranges for both lines extend from line centre to approximately $-420$\,km\,s$^{-1}$ and $+480$\,km\,s$^{-1}$.
The spatial separation and PA of the centroids of these two channels are $\sim1.5\arcsec$ and 21$\degr$. 
 %Blue-wing and red-wing indicate the integrated blue channels and red channels of molecular line. 
The absorption is clearly visible in the central region for both molecules.
In particular, HCO$^+$ has stronger absorption than HCN. 
While this means we cannot reliably fit a Gaussian to the HCO$^+$ emission, it is still possible to do so on the HCN map.
The resulting FWHM is $2.94\arcsec \times 1.44\arcsec$ and PA is 167$\degr$ (Table~\ref{tab:emi_info}).

Panel (a) of Figure~\ref{fig:ngc3079spec} shows the integrated spectrum of NGC\,3079.
Although a clear P-Cygni line profile is apparent for the HCN\,(1-0) and HCO$^+$\,(1-0) line due to the strong emission of these lines, blue-shifted absorption with two or more components dominates the profile of other lines where an emission component is weak or absent. 
These include H$^{13}$CN\,(1-0) at 86.34\,GHz, H$^{13}$CO$^+$\,(1-0) at 86.75\,GHz, SiO\,(2-1) at 86.85\,GHz, and HN$^{13}$C\,(1-0) at 87.09\,GHz (rest frequencies).
The narrow absorption peak closest to systemic velocity is strongly absorbed, reaching maximum depths of 87\%, 56\%, 71\%, and 42\% of the continuum level respectively. 
The absorption profiles show a second narrow peak with a blue-shifted tail (that may itself contain several sub-components) that reaches to -350\,km\,s$^{-1}$.
The absorption that we have identified as C$_2$H at 87.3--87.4\,GHz is at relatively low signal-to-noise and is likely to include features from several of the hyperfine transitions \citep{got83,Schoier2005}.
We therefore exclude this from our analysis.

Panel (b) of Figure~\ref{fig:ngc3079spec} shows the double peaked absorption of H$^{13}$CN\,(1-0).
While there may be substructure within the full velocity extent of the absorption, for our analysis, and discussion of the implications, in Section~\ref{sec:cd3079} we consider this as only two features.
The reason is that we prefer to interpret the absorption in the context of its physical origin, and we argue in Section~\ref{sec:cd3079} that the narrow peak closer to systemic is due to absorption by the nuclear disk, while the more blue-shifted component, as well as its high velocity tail, originates in absorption by outflowing gas clouds.
We note also that the absorption profile of the H$^{13}$CN\,(1-0) line is very different to that of HCN\,(1-0).
This is for two reasons: 
it is not saturated while, as discussed below, the HCN\,(1-0) absorption is strongly saturated (and as a result one cannot use the H$^{13}$CN\,(1-0) profile to correct for the HCN\,(1-0) absorption); 
and because any associated line emission is very weak, the absorption profile is not partially filled, as is the case for the HCN\,(1-0) line.
These issues are discussed in detail below in Section~\ref{sec:abscorr}.

There are few reports of HCN\,(1-0) or HCO$^+$\,(1-0) absorption in the literature.
This could be because most mm spectra of galaxies are still from single dish rather than interferometric observations and, as we discuss below in Section~\ref{subsub:Compare_ngc3079}, extended low intensity emission will tend to fill in any absorption of a weak continuum.
The two cases of which we are aware, blue-shifted HCO$^+$\,(4-3) absorption in Arp\,220 \citep{sak09} and red-shifted HCN\,(3-2) and HCO$^+$\,(3-2) absorption in IC\,860 \citep{aal15}, were both obtained with sub-arsec resolution interferometric observations.

\subsubsection{Correction for absorption}
\label{sec:abscorr}

The observed (i.e. without correcting for the absorption) flux of HCN\,(1-0) is 8.20\,Jy\,km\,s$^{-1}$ and of HCO$^+$\,(1-0) is 4.78\,Jy\,km\,s$^{-1}$, both in 5\arcsec\ apertures to cover the full extent of the observed emission.
In order to estimate a correction for the absorption, we fit a Gaussian function to the emission profile, including an additional constraint that the intrinsic absorption as measured from the continuum level cannot go below zero.
This approach is motivated under the assumption that the absorption and emission are effectively independent: the absorption at any given velocity occurs over a very small area (defined by the size and location of continuum sources) while the emission is integrated over the full extent of that isovelocity contour across the molecular disk (note that we did not require that the centre of the symmetric emission should be exactly at the systemic velocity of molecular line because the line emission may not be symmetric about the centre of the rotating disk), and partially fills the absorption.
The best fitting profile is shown as a solid green line in the left panel of Figure~\ref{fig:3079profiles}, where the black solid line traces the observed HCN\,(1-0) profile.
The difference between these is represented by the dashed green line, which traces the intrinsic absorption profile from the continuum level.
Its truncated shape indicates that the HCN\,(1-0) line is strongly saturated.
The strong red wing of the observed emission profile, and the constraint on the depth of the intrinsic absorption, result in a fit whose peak is offset by $\sim100$\,km\,s$^{-1}$ from systemic and which under-estimates the blue wing of the emission line profile.
In addition, the continuum absorption lacks the blue wing seen in the H$^{13}$CN profile.
To avoid these discrepancies we have made a second reconstruction, which is shown in the right panel of Figure~\ref{fig:3079profiles}.
For this, we have added the requirement that the intrinsic Gaussian line profile (dotted blue line) should be centered at the systemic velocity.
We find that, because the continuum absorption (dashed blue line) cannot go below zero, the reconstructed line profile (solid blue line) cannot be Gaussian -- noting that the sum of these two must match the observed profile.
Instead, in comparison to the Gaussian, it shows absorption that is roughly symmetric and centered approximately at the systemic velocity.
We interpret this to mean that the HCN\,(1-0) line is self-absorbed.
Self-absorption in an HCN\,(1-0) line has been reported in IC\,860 and Arp\,220\,W by \cite{aal15} in the context of the compact obscured nuclei of luminous and ultraluminous infrared galaxies.
It has also been reported in the CO\,2-1 and CO\,6-5 lines in Arp\,220 by \cite{eng11} and \cite{ran15} respectively.
In the case of NGC\,3079 we suggest that it is occuring because the galaxy is inclined so close to edge on.
As such, there is plenty of cool gas in the disk of the galaxy along the line of sight, that could absorb the line emission.
From the depth of the self-absorption we estimate that a total column of $N_{HCN}=1.4\times10^{14}$\,cm$^{-2}$ is needed to produce it, corresponding to $N_H\sim10^{21}$\,cm$^{-2}$ if we adopt the same abundance as in Section~\ref{sec:cd3079} or as much as a few $\times10^{22}$\,cm$^{-2}$ for a more typical lower HCN abundance.
The absorption is at the systemic velocity because the disk motion along the line of sight to the nucleus is across the plane of the sky.

Based on our reconstruction, the flux of the self-absorbed HCN\,(1-0) line is 16.1\,Jy\,km\,s$^{-1}$ (nearly a factor 2 greater than the observed flux) while the total flux of the intrinsic Gaussian is 18.7\,Jy\,km\,s$^{-1}$.
A similar analysis of the HCO$^+$\,(1-0) line confirms that it does suffer more from absorption: our estimate of the self-absorbed flux is 13.3\,Jy\,km\,s$^{-1}$ (nearly a factor 3 greater than that measured directly) and the intrinsic flux in the Gaussian is 19.5\,Jy\,km\,s$^{-1}$.
The FWHM of the intrinsic Gaussian profile is 360\,km\,s$^{-1}$, which we have included in Table~\ref{tab:emi_info}.

\subsubsection{Comparison with single dish measurements}
\label{subsub:Compare_ngc3079}

A spectrum of NGC\,3079 including the lines discussed above was presented previously by \cite{Costagliola2011}.
This was based on data from the IRAM 30\,m telescope, with a beam size of 29\arcsec\ at 88\,GHz.
%While the HCN\,(1-0) and HCO$^+$\,(1-0) lines both show a double-peaked profile -- indicative of either continuum absorption or self-absorption -- their fluxes of 24.6 and 27.7\,Jy\,km\,s$^{-1}$ are several times greater than observed with our interferometric beam, and the line widths of 500\,km\,s$^{-1}$ FWHM are significantly broader.
While the HCN\,(1-0) and HCO$^+$\,(1-0) lines both show a double-peaked profile -- which can be generally interpreted as disk rotation or be indicative of either continuum absorption or self-absorption in the centre -- their fluxes of 24.6 and 27.7\,Jy\,km\,s$^{-1}$ are several times greater than observed with our interferometric beam, and the line widths of 500\,km\,s$^{-1}$ FWHM are significantly broader.
They also detect a strong blend of C$_2$H emission at about 87\,GHz which we detect only weakly.
The difference in line fluxes suggests that there is significant, but low intensity, emission from these lines on scales greater than a few arcsec.
As is apparent from our data, as one integrates the flux within larger apertures, the absorption below the continuum level is filled in by the additional line emission included in the aperture, and so the line appears broader.
It is likely that in the single dish measurement, continuum absorption has only a minor impact, and the double peaked profile is likely to be due to self-absorption (we note that the dip between the peaks is to a similar depth as our reconstructed self-absorbed profile).
If one does not correct for this, the line width will be overestimated.
For the interferometric spectrum in Figure~\ref{fig:3079profiles}, we can estimate the effective FWHM of the self-absorped profile by fitting a single Gaussian to the double-peak profile.
Approximating the profile in this way yields a FWHM of 450--470\,km\,s$^{-1}$, similar to the 500\,km\,s$^{-1}$ width reported by \cite{Costagliola2011} but significantly more than the intrinsic FWHM of 360\,km\,s$^{-1}$.
Similarly, once one corrects for the continuum absorption, the ratio of the HCN\,(1-0) and HCO$^+$\,(1-0) line fluxes is similar to that reported by \cite{Costagliola2011}.
We conclude that while our spectrum looks rather different from the single dish spectrum, all the differences can be understood as a result of extended low intensity emission and the aperture dependent impact of continuum absorption on the observed profile.

\subsection{NGC\,6764}

The 3\,mm continuum map is shown in the bottom left panel of Figure~\ref{fig:conhcnhco}.
Its FWHM of $1.04\arcsec \times 0.82\arcsec$ and PA of $8\degr$ indicate it is spatially unresolved. 
Within a 3\arcsec aperture, the continuum flux density is 0.6\,mJy.

\begin{figure}
\begin{center}
      \includegraphics[width=70mm]{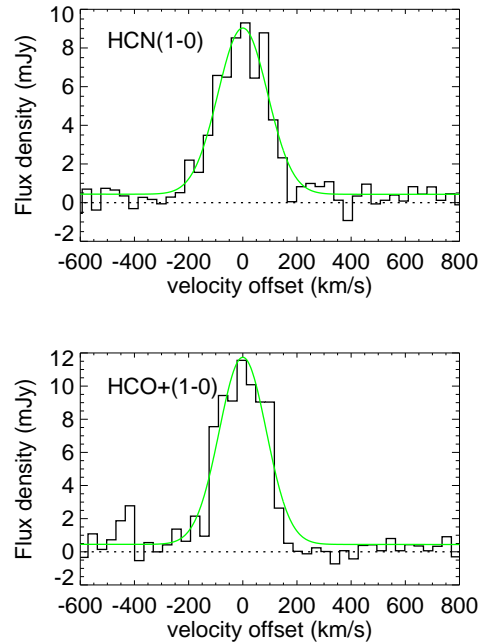}      
         \caption{Spatially integrated spectrum for NGC\,6764 showing the HCN\,(1-0) (top panel) and HCO$^+$\,(1-0) (bottom panel) emission lines. 
The continuum has not been subtracted. 
In each case the best fitting Gaussian function is represented by the green line. 
The line profiles are very similar, showing a large velocity dispersion with a mean FWHM of 209\,km \,s$^{-1}$.}
\label{fig:ngc6764spec}
\end{center} 
\end{figure}

Figure~\ref{fig:conhcnhco} also shows the integrated HCN\,(1-0) and HCO$^+$\,(1-0) maps in the bottom middle and right panels, and the measurements and the line fluxes are summarised in Table~\ref{tab:emi_info}.
The HCN\,(1-0) emission has an extent of $1.7\arcsec\times1.3\arcsec$ at PA $-54\degr$, similar to the $1.7\arcsec\times1.2\arcsec$ at PA $-66\degr$ of the HCO$^+$\,(1-0) line.
The red and blue channels of emission (from line centre to $-435$\,km\,s$^{-1}$ and $+465$\,km\,s$^{-1}$ for HCN; to $-335$\,km\,s$^{-1}$ and $+365$\,km\,s$^{-1}$ for HCO$^+$) also have the same separation of 0.82\arcsec\ at PA $-87\degr$.
Figure~\ref{fig:ngc6764spec} shows integrated spectra covering the two line profiles including a constant continuum level. 
Their FHWMs are 214\,km\,s$^{-1}$ and 203\,km\,s$^{-1}$ respectively.
Taken together, these measurements indicate that the distributions and kinematics of the two lines are very similar, and hence that they originate in the same region.
This means that when constructing a simple dynamical model, as discussed in Sec.~\ref{sec:model_kinematics}, we can constrain it using both lines, which provides an additional robustness against limitations of signal-to-noise or line specific peculiarities.

\subsection{NGC\,5033}
\label{sec:Rngc5033}

\label{sec:Rngc6764}

\begin{figure}
\begin{center}
      \includegraphics[width=70mm]{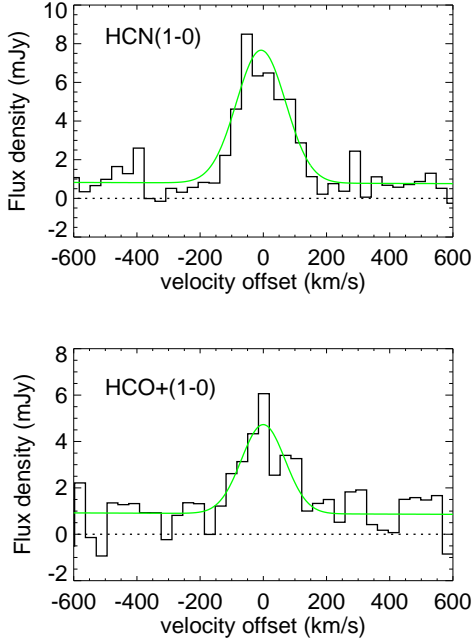}      
         \caption{Spatially integrated spectrum for NGC\,5033 showing the HCN\,(1-0) (top panel) and HCO$^+$\,(1-0) (bottom panel) emission lines. 
The continuum has not been subtracted. 
In each case the best fitting Gaussian function is represented by the green line. 
%In contrast to NGC\,6764, both lines are reasonably well centered at the same (systemic) velocity.
And the line profiles are very similar, showing a large velocity dispersion with a mean FWHM of 183\,km \,s$^{-1}$.}
\label{fig:ngc5033spec}
\end{center} 
\end{figure}

The 3\,mm continuum for NGC\,5033 is presented in the top left panel of Figure~\ref{fig:conhcnhco}, and has a flux density within a 3\arcsec\ aperture of 0.7\,mJy.
A Gaussian fit gives a size of $1.01\arcsec \times 0.79\arcsec$ at PA 41$\degr$. 
As for the other 2 galaxies, this is consistent with the continuum being spatially unresolved.

The HCN\,(1-0) and HCO$^+$\,(1-0) maps are shown in the top middle and right panels of Figure~\ref{fig:conhcnhco} respectively. 
The measured sizes are $2.0\arcsec \times1.1\arcsec$ at PA $-17\degr$ and $1.7\arcsec \times1.2\arcsec$ at PA $52\degr$ respectively. 
We note that the latter case, HCO$^+$\,(1-0) has a large uncertainty which we attribute to the weaker HCO$^+$\,(1-0) line flux.
As before, the blue and red contours represent the line emission from blue and red channels (from line centre to $-435$\,km\,s$^{-1}$ and $+465$\,km\,s$^{-1}$ for HCN; to $-335$\,km\,s$^{-1}$ and $+365$\,km\,s$^{-1}$ for HCO$^+$).
Their separations are $1.1\arcsec$ and $1.0\arcsec$ respectively.
The PA for HCN\,(1-0) is $-17\degr$; the PA for HCO$^+$\,(1-0) differs from this but has a large uncertainty because of the low signal-to-noise in the blue channels, and is still consistent with the HCN\,(1-0).
Figure~\ref{fig:ngc5033spec} shows spectra extracted within a $3\arcsec$ aperture.
The flux of HCN\,(1-0) is 1.2\,Jy\,km\,s$^{-1}$ and of HCO$^+$\,(1-0) is 0.71\,Jy\,km\,s$^{-1}$. 
The two lines have similar width, with a mean FWHM of 183\,km\,s$^{-1}$. 
Again, we find that the distributions and kinematics of the two lines are similar, and hence conclude that they originate from the same region.

\section{Modelling the kinematics}
\label{sec:model_kinematics}

The description above of the observed characteristics shows the FWHM of the line is typically 200\,km\,s$^{-1}$.
Comparable line widths for the HCN\,(1-0) line were reported previously for the central hundred parsecs of several AGN by \cite{Sani2012}.
Our aim in this Section is to understand whether the line widths can be accounted for by a thin rotating disk, or whether it implies there is a significant intrinsic dispersion associated with the molecular disk of dense gas.
To do so we create a simple dynamical model that can match the observed characteristics of the molecular line emission, using the same methodology as \cite{Sani2012}.

To constrain our model we use four of the observed molecular gas properties listed in Table~\ref{tab:emi_info}:
(1) major axis FWHM, (2) minor axis FWHM, (3) separation between the centers of the blue and red channel maps, and (4) spectral FWHM of the spatially integrated line.
The inclination and PA of the disk are fixed input parameters, as described below.
We assume that the line traces a rotating disk and model the observed properties by applying the IDL code DYSMAL (Dynamical Simulation and Modelling Algorithm, described in \citealt{Davies2011}). 
The main purpose of this code is to quantify the impact of spectral and spatial beam smearing on an axisymmetric rotating disk.
In doing so it allows us to infer the intrinsic kinematics of the disk from the observed properties. 
%We exclude NGC\,3079 from this analysis on account of its strong absorption, which influences the accuracy of morphological and spectral properties.
On the other hand, we include NGC\,7469 using data presented in \cite{Davies2004}.
While the HCN\,(1-0) observations for that object have not been modelled in this way, the 0.7\arcsec\ resolution 1\,mm CO\,(2-1) and 0.09\arcsec\ resolution 2.12\,$\mu$m H$_2$ 1-0\,S(1) data enabled those authors to derive a detailed mass distribution based on the combined dynamics at their different resolutions.
This object therefore allows us to directly test whether our simple dynamical model based on a single Gaussian mass distribution can be considered a working approximation in the context of disk size scale, dynamical mass, and ability to distinguish between thin and thick geometries.

\subsection{Kinematic Modelling Procedure}

The kinematic modelling procedure we use is the same as described in \cite{Sani2012}.
However, in addition, we explore the impact of the initial conditions on the convergence of the minimisation routine to ensure that we have reached the global rather than local minimum.
To do so we generate a set of random values within broad but restricted ranges for the disk size, disk thickness, and a mass scaling ($M_{scale}$, which is simply a way to set the amplitude of the rotation curve, and represents the mass supported by ordered circular orbits in the disk plane).
We then run a minimisation starting from these values.
At each iteration, it uses DYSMAL to simulate how a disk model with the given properties would appear when convolved with the spatial beam and spectral resolution of our data.
The model is based on a Gaussian distribution, the FWHM of which is equal to the given disk size ($R$ is half of this FWHM).
The shape and amplitude of the rotation curve is derived from this in combination with the mass scaling $M_{scale}$.
Finally, the vertical profile of the disk is defined by a Gaussian distribution, the FWHM of which is equal to the given disk thickness ($H$ is half of this FWHM).
The thickness has an observational impact both on the spatial distribution if the disk is inclined, and also on the velocity dispersion which we calculate as $\sigma = vH/R$ and assume is isotropic.
The line of sight velocity distribution is derived for each point through the inclined disk model; % (see Section~\ref{sec:Obs} for the adopted inclinations);
and, after applying appropriate beam smearing, used to generate an output cube with 2 spatial axes and 1 velocity axis. 
The output data cube is analysed in the same way as the real data, extracting the four properties listed above and comparing them with those extracted from the observations.
The minimisation routine iteratively converges on a set of disk parameters that lead to a best match of the observed properties.
After repeating this process 100 times with different initial parameters, we obtain 100 sets of the best-fit model output parameters and their corresponding $\chi^{2}$.
We select the 50 sets with the lowest $\chi^{2}$ and use these to estimate the mean for each model parameter (note that we use the statistical mean rather than the set of parameters from any single minimisation), and list them in Table~\ref{tab:model_output}. 
In a final step, we input the mean values back into DYSMAL one more time and retrieve the simulated output `observable' properties, which are also listed in Table~\ref{tab:model_output}.
Together with these output parameters of the model, we list the input parameters (disk size, disk thickness, and scaling M$_{scale}$).
To constrain the input parameter uncertainties, we estimate the input parameter distribution (either side of the best-fit values) that can satisfy the output observable properties within their uncertainties. This is similar in principle to the Markov chain Monte Carlo method, but simplified since we do not require to know the full probability distributions for the input parameters.
We list the $\pm$1$\sigma$ uncertainties for input parameters in Table~\ref{tab:model_output}. 
We summarise the associated kinematics of the best-fit rotating disk at a radius $R$ in Table~\ref{tab:model_int}, together with the enclosed dynamical mass $M_{dyn}$ which we estimate as $(v^2+3\sigma^2)R/G$ (as discussed in \citealt{Davies2007} and \citealt{Sani2012}).

\begin{table*}
  \caption{Summary of simulated disk models: \newline 
(1) Source name; (2) Molecule; (3) Note; (4) Disk FWHM size in the disk plane; (5) Disk FWHM thickness (height); (6) Scaling $M_{scale}$ (which sets the amplitude of the rotation curve, and represents the mass supported by ordered circular orbits in the disk plane); (7)-(10) are measured with the same method as columns (4), (5), (7), and (9) of Table~\ref{tab:emi_info}; (11) reduced $\chi^2$ of the fit.}
    \begin{tabular}{*{11}{c}} \hline         
    Source & Molecule & Note$^{(a)}$ & \multicolumn{3}{c}{Mean value of best-fit} & \multicolumn{3}{c}{4 simulated properties} \\
          &   &   & \multicolumn{3}{c}{model parameters} & \multicolumn{3}{c}{from DYSMAL} & $\chi^2_{red}$ \\
       & &  & size & thickness & log(M$_{scale}$)  & Maj. $\times$ min. & b/r sep. & Line width  & \\
       & & & (arcsec) & (arcsec) & (M$_{\sun}$) &  axis (arcsec$^2$) & (arcsec) & (km s$^{-1}$) & \\ 
\hline \hline
NGC\,7469 & HCN & Complex & 2.41$^{+0.01}_{-0.05}$ & 0.43$^{+0.01}_{-0.03}$ & 9.95$^{(b)}$         & 3.15 $\times$ 2.69 & 1.22 & 242 & 6.3 \\
          & HCN & Simple  & 2.56$^{+0.03}_{-0.09}$ & 0.62$^{+0.03}_{-0.06}$ & 9.42$^{+0.02}_{-0.02}$ & 3.27 $\times$ 2.84 & 1.15 & 235 & 4.5 \\
NGC\,6764 & HCN & -       & 1.35$^{+0.03}_{-0.10}$ & 0.25$^{+0.01}_{-0.14}$ & 8.66$^{+0.05}_{-0.07}$ & 1.71 $\times$ 1.21 & 0.74 & 203 & 2.2 \\
          & HCO$^+$  & -  & 1.43$^{+0.03}_{-0.10}$ & 0.26$^{+0.01}_{-0.15}$ & 8.75$^{+0.06}_{-0.04}$ & 1.77 $\times$ 1.23 & 0.79 & 217 & 6.6 \\
NGC\,5033 & HCN & -       & 1.70$^{+0.08}_{-0.15}$ & 0.01$^{+0.06}_{-0.00}$ & 8.47$^{+0.05}_{-0.15}$ & 1.94 $\times$ 1.17 & 0.98 & 182 & 0.7 \\
          & HCO$^+$ & -   & 1.54$^{+0.09}_{-0.14}$ & 0.19$^{+0.08}_{-0.12}$ & 8.37$^{+0.05}_{-0.15}$ & 1.85 $\times$ 1.18  & 0.88 & 190 & 0.1 \\
NGC\,3079 & HCN & -       & 2.66$^{+0.02}_{-0.03}$ & 0.45$^{+0.01}_{-0.05}$ & 9.30$^{+0.03}_{-0.06}$ &  2.92 $\times$ 1.43 & 1.51 & 378 & 0.9 \\
\hline
    \end{tabular} 
    \\ 
 Note: the uncertainties given are derived as joint uncertainties and hence implicitly take into account possible partial correlations between parameters.\\
$^{(a)}$For NGC\,7469 'complex' refers to the multi-component dynamical model from \cite{Davies2004}, 'simple' to the single Gaussian representation for the mass distribution. \\
 $^{(b)}$ Fixed to $M_{scale}=9 \times 10^{9}$\,M$_\odot$ as given by  \cite{Davies2004}. 
    \label{tab:model_output}   \\
\end{table*}

\begin{table*}
 %\begin{center}
  \caption{Intrinsic kinematics of the modelled rotating disks: \newline
(1) Source name; (2) Molecule; (3) Radius (half of the FWHM given in Table~\ref{tab:model_output}); (4) Rotational velocity at $R$; (5) Velocity dispersion at $R$; (6) Enclosed dynamical mass estimated as $(v^2+3\sigma^2)R/G$.}
    \begin{tabular}{llrrrr} \hline               
     Source & molecule & R  & v   &  $\sigma$ & log(M$_{dyn}$) \\
        & & (pc) &  (km s$^{-1}$) & (km s$^{-1}$) & (M$_{\sun}$) \\ 
  \hline \hline
NGC\,7469  &HCN$^{(a)}$  & 356 & 131 & 47 & 9.30  \\
           &HCN$^{(b)}$  & 377 & 128 & 62 & 9.39  \\
NGC\,6764  &HCN         & 101 & 102 &  38 & 8.54  \\
           &HCO$^+$     & 108 & 110 &  40 & 8.63  \\
NGC\,5033  &HCN         &  62 & 104 & 1.4 & 8.20  \\
           &HCO$^+$     &  56 &  98 &  25 & 8.18  \\
NGC\,3079  &HCN         & 112 & 204 & 69  & 9.16 \\ 
\hline
    \end{tabular} 
    \\    
    %\end{center}
    %R = (disk width)/2 * (pc/")
     $^{(a)}$Complex model with multiple components; $^{(b)}$ Simple model with 1 component.
     \label{tab:model_int}    \\
\end{table*}

\subsection{NGC\,7469}

Although the interferometric beam size at 3\,mm is only $\sim1$\arcsec, this is still relatively large compared to the scale of the nuclear structures.
As a result, models are necessarily very simple and we have used a single Gaussian mass distribution to define the rotation curve and emission distribution.
Since we are limited to using a very simple model, we use NGC\,7469, which has been already successfully modeled with a complex mass distribution, to assess whether our approach is applicable.
\citet{Davies2004} constructed a single axisymmetric mass model for NGC\,7469 comprising a broad disk, a ring and an extended nucleus to interpret the observations of CO\,(2-1) and the K-band H$_{2}$ 1-0\,S(1) line at different resolutions. 
The details of this mass model are summarized in their Table~2, and we adopt it for our analysis, calling it the `complex model'.
While the mass distribution represented by these 3 components is fixed, we represent the HCN\,(1-0) line emission distribution by a single independent component (i.e. the kinematics of the line are defined by the fixed mass distribution, but the distribution of the line emission does not necessarily follow that of the mass).
One additional assumption we make is that all the components share the same thickness.
This is a free parameter in the model, as is the size scale of the luminosity distribution, while the mass scale (as before, this represents only the mass supported by ordered circular orbits in the disk plane) is fixed to $M_{scale}=9 \times 10^{9}$\,M$_\odot$ as derived by \cite{Davies2004}.
When applying our kinematics modelling procedure as described above, the best-fit values for the disk size and disk thickness are 2.41$\arcsec$ and 0.43$\arcsec$ respectively.
And, as summarised in Table~\ref{tab:model_int}, accounting for rotation and dispersion, we estimate the dynamical mass to be $\log{M_{dyn}[M_\odot]} = 9.30$.
In comparison, when using a single Gaussian profile to define the mass distribution (hereafter called the `simple model'), we leave the disk size, disk thickness, and mass scale as three free parameters.
We then derive best-fit mean values for disk size and thickness to be 2.56$\arcsec$ and 0.62$\arcsec$, and estimate $\log{M_{dyn}[M_\odot]} = 9.39$.
There are two key results here.
The first is that the disk size and thickness are consistent between the two models, as is $M_{dyn}$.
The second is that the model implies the disk is thick, with an intrinsic $\sigma/v \sim 0.4$.
While the dispersion we derive of 50--60\,km\,s $^{-1}$ is larger than the 30\,km\,s$^{-1}$ seen directly in the higher spatial resolution CO\,(2-1) data by \cite{Davies2004}, both would lead to the same conclusion in the context of distinguishing between a thick disk and a thin disk: that a significant intrinsic dispersion is required, and a thin disk is ruled out.

Thus we can estimate the intrinsic kinematics, and in particular the dispersion (thickening), for the disk using a simple approximation to the mass distribution.
In the rest of this Section, we make use of the simple model to derive the intrinsic kinematics also for NGC\,6764 and NGC\,5033. 

\begin{figure*}
\begin{center}
      \includegraphics[width=120mm]{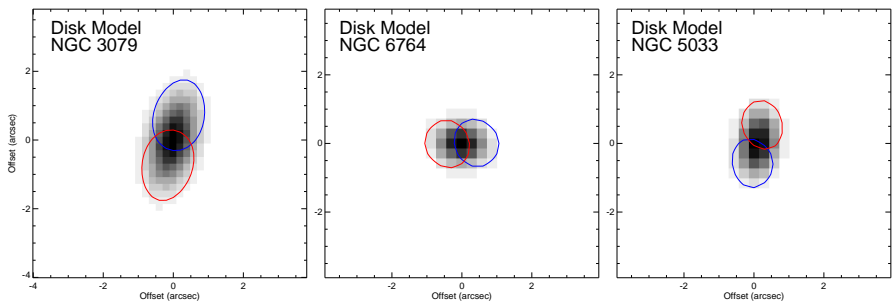}      
      \caption{Simulated disk models for the 3 sources with new observations, convolved with the appropriate beam. As for Figure~\ref{fig:conhcnhco}, the greyscale image shows the emission of the modelled disk while the blue and red contours represent the distribution of integrated blue and red channels of the emission lines (at 50\% of the peak intensity). We show the model for only one line, but it is a good representation for both HCN\,(1-0) and HCO$^+$\,(1-0). A PV diagram extracted from the model for NGC\,3079 is shown in Figure~\ref{fig:PVdiagram}.}
\label{fig:diskmodels}
\end{center} 
\end{figure*}

\subsection{NGC\,6764}

When modelling NGC\,6764, we assume that the nuclear disk and galactic disk have the same inclination, which we set to $62\degr$ \citep{Leon2007}.
And we note that the PA of the line emission, as well as its velocity gradient, are approximately perpendicular to the parsec-scale core-jet structure (PA $\sim 25\degr$, \citealt{Kharb2010}). 
We caution that, while the velocity gradient we see at radii up to 1\arcsec\ is consistent with that of the CO\,(2-1) line up to radii of 5\arcsec\ \citep{Leon2007}, our model focusses on the smaller scale and may not represent the properties of the rotating disk at $\ga$200\,pc, which might be perturbed by streaming motions associated with the bar.

We perform the kinematics modelling separately on each of the two lines, which independently yield essentially the same solution:
a disk with a size of $\sim1.4\arcsec$ and a thickness of $\sim0.25\arcsec$.
We conclude that the observational constraints require a thick disk model to explain the line emission distribution and kinematics, and in particular the large velocity width of the molecular lines.
In our assumption of hydrostatic equilibrium, the disk thickness of the model corresponds to an intrinsic $\sigma/v \sim 0.37$.

\subsection{NGC\,5033}

For NGC\,5033, based on the kinematical parameters derived from the CO\,(1-0) velocity field \citep{Kohno2003}, we fix the inclination to $66\degr$.
We note also that the PA on those larger scales matches what we have measured on 1\arcsec\ scales for the HCN\,(1-0) and HCO$^+$\,(1-0) lines, and is approximately perpendicular to the synchrotron radiation from the core-jet structure, which is oriented east-west \citep{Perez2007}.

Our modelling results differ for the two lines.
The characteristics of the HCN\,(1-0) line are well matched by a very thin disk with a size of $1.7\arcsec$.
In contrast, the HCO$^+$\,(1-0) emission is better matched by a thicker disk with a size of $1.54\arcsec$ and a thickness of $0.19\arcsec$.
However, due to the low signal-to-noise of the HCO$^+$\,(1-0) line emission, the observable properties have large uncertainties and so the model is poorly constrained, as reflected in the very small $\chi^2$ for this fit.
For this object there is no convincing evidence for a thick disk.
We conclude that our kinematic analysis favours the thin disk solution, for which the corresponding intrinsic $\sigma/v$ ratio is $<0.1$.

\begin{figure}
\begin{center}
      \includegraphics[width=75mm]{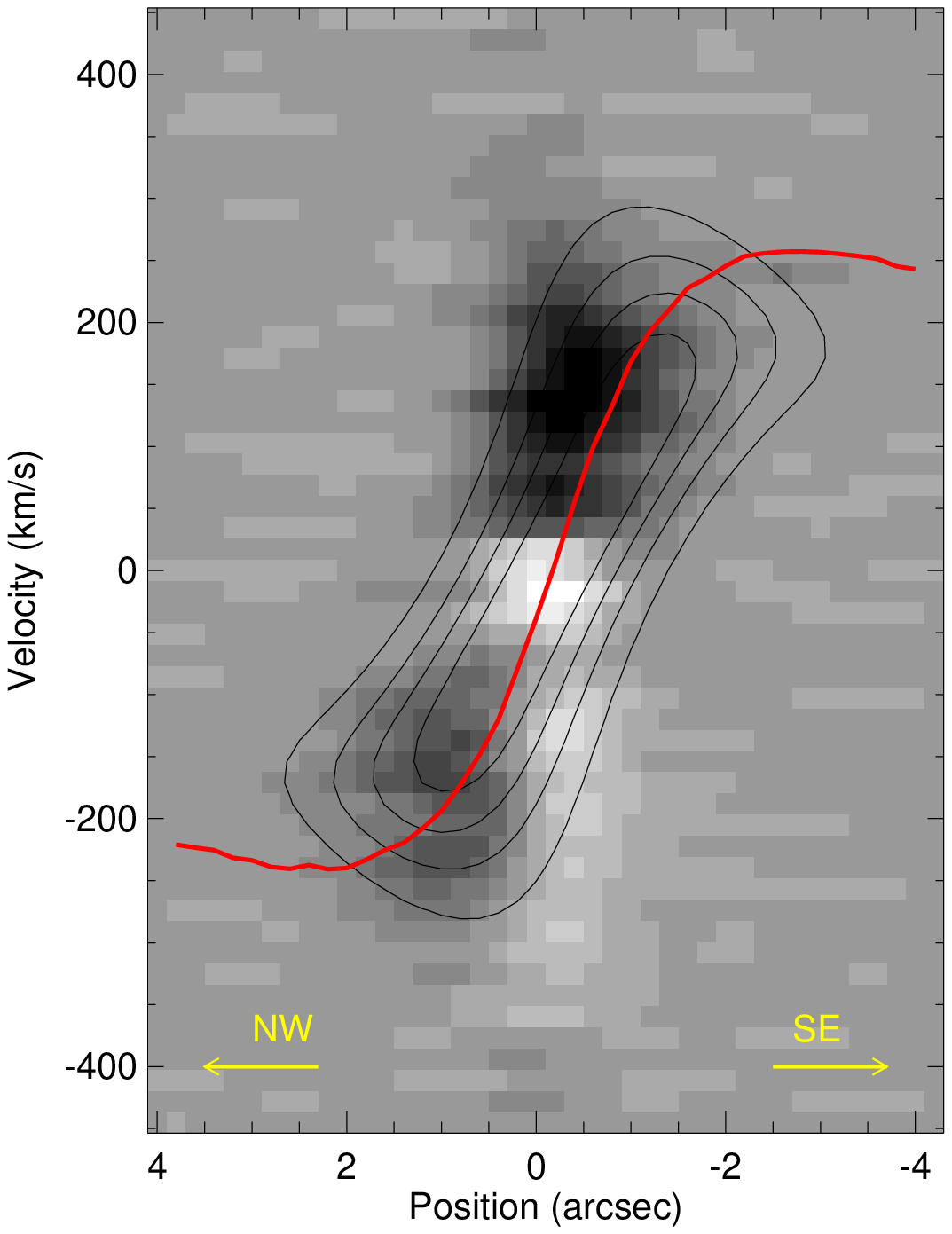}      
\caption{The observed PV diagram (grey scale) for the HCN\,(1-0) line in NGC\,3079, extracted along the major axis (slit $1\arcsec$ wide at a PA of $165\degr$).
The beam-convolved PV diagram from our axisymmetric dynamical model is overplotted as black contours ($20\%$, $40\%$, $60\%$, and $80\%$ of the peak intensity).
The red line traces the rotation curve derived from it using the iteration method \citep{Takamiya2002}.
Positive and negative positions indicate the NW and SE direction respectively.}
\label{fig:PVdiagram}
\end{center} 
\end{figure}

\subsection{NGC\,3079}

Despite the absorption in the molecular emission lines, we still apply the kinematics modelling to HCN\,(1-0);
however, we exclude HCO$^+$\,(1-0) since the effect of the absorption is too severe.
According to the kinematical parameters derived from the CO\,(1-0) velocity field \citep{Koda2002}, we fix the nuclear disk inclination to $77\degr$ and PA to $165\degr$.
In addition, to reduce any bias caused by the impact of the absorption in setting the amplitude of the rotation curve, we also fix $M_{scale}=2 \times 10^{9}$\,M$_\odot$ (at 3\arcsec) based on the value derived by these authors from their kinematical modelling of the CO\,(2-1) line of their `core' component, which we identify as the component traced by the HCN\,(1-0) and HCO$^+$\,(1-0) lines.
The best-fit mean values for the disk size and thickness are $2.66\arcsec$ and $0.45\arcsec$.
The associated dispersion is just under 70\,km\,s$^{-1}$, consistent with the 60\,km\,s$^{-1}$ of \cite{Koda2002}, and confirms that the dense clouds exhibit significant intrinsic random motions.
The observed line emission distributions and kinematics need the thick disk geometrical structure, the corresponding intrinsic $\sigma/v$ ratio is $\sim 0.33$.

Since the emission lines are better spatially resolved in NGC\,3079 than the other targets, we plot a position-velocity (PV) diagram in Figure~\ref{fig:PVdiagram} extracted along the major axis.
The greyscale image shows that emission as darker colours and the absorption in lighter colours.
Superimposed are contours tracing our axisymmetric dynamical model.
Given the simplicity of this model, it traces the emission very well, although the line emission does shows deviations from axisymmetry (in particular, the south eastern side shows indications for red-shifted emission at $V>200$\,km\,s$^{-1}$ at a radial offset of $\la0.5\arcsec$). 
We have used the iteration method \citep{Takamiya2002,Sofue2001} to derive the central rotation curve from the beam-convolved model, and overplotted this on the figure.
The model compares well with Figure~15 of \cite{Koda2002}, where the spatial resolution was slightly poorer than ours, but which was not hampered by absorption.

\begin{table}
 \caption{Summary of the kinematic modelling for the combined sample which includes the 4 Seyferts analysed here and 4 Seyferts from \citet{Sani2012}. \newline
(1) Source name; (2) HCN\,(1-0) Luminosity (3) Radius; (4) Rotational velocity at $R$; (5) Velocity dispersion at $R$; (6) Enclosed dynamical mass.}
    \begin{tabular}{lrrrrr} \hline               
     Source  & L$_{HCN}$ & R    & v             &  $\sigma$      & M$_{dyn}$ \\
             &          & (pc) & (km\,s$^{-1}$) & (km\,s$^{-1}$) & (M$_\odot$) \\ 
  \hline \hline
NGC\,7469$^{a}$ & 7.87     & 356 & 131 & 47 & 9.30 \\
NGC\,6764$^{b}$ &  6.90    & 101 & 102 & 38 & 8.54 \\
NGC\,5033$^{b}$ &  6.36    &  62 & 104 &  1 & 8.20 \\
NGC\,3079      &  7.49    & 112 & 204 & 69 & 9.16 \\
NGC\,2273       & 6.86     & 97 &  83 & 33 & 8.36 \\
NGC\,3227       & 6.52     & 23 & 105 & 42 & 7.98 \\
NGC\,4051       & 5.88     & 22 &  47 & 19 & 7.23 \\
NGC\,6951       & 6.36     & 45 &  84 & 34 & 8.04 \\ \hline
    \end{tabular} 
\label{tab:combined}    \\
$^{(a)}$From the complex model.\\
$^{(b)}$We have adopted the model based on the HCN\,(1-0) line: for NGC\,6764 it provides a better fit; for NGC\,5033 it provides a better constrained fit.
\end{table}

\subsection{Kinematics Summary}

For three of the four galaxies we have modelled, the nuclear emission line characteristics require a thick disk with $\sigma/v \ga 0.3$.
The validity of our results is demonstrated for 2 objects.
For NGC\,7469, matching the kinematics of higher resolution CO\,(2-1) and H$_2$ 1-0\,S(1) data to a detailed mass model, \cite{Davies2004} noted that the dispersion of the CO\,(2-1) line had to be a relatively high 30\,km\,s$^{-1}$.
For NGC\,3079, focussing on the `core' component seen in CO\,(2-1) data, \cite{Koda2002} required a dispersion of 60\,km\,s$^{-1}$ to model the kinematics.

\citet{Sani2012} performed a similar kinematic analysis of the HCN\,(1-0) line for four other nearby Seyfert galaxies, finding $\sigma/v$ in the range 0.3--0.5 for all of them.
We combine these with our kinematic analysis above and summarize the combined modelling results in Table~\ref{tab:combined}.
We find that in 7 out of 8 nearby Seyferts, $\sigma/v \ga 0.3$.
Only for NGC\,5033 do we find evidence that the disk is thin, with  $\sigma/v \la 0.1$.
We conclude that thick molecular disks are a common occurrence in the central $\sim100$\,pc of Seyfert galaxies.

\section{Column density through the inner disk of NGC\,3079}
\label{sec:cd3079}

In this section, we use the depth of the absorption features in the spectrum of NGC\,3079 to make a direct estimate of the gas column density of the disk, and discuss the origin of the absorption features.

We already noted in Section~\ref{sec:Rngc3079} that the H$^{12}$CN\,(1-0) absorption is saturated, making it difficult to use reliably to derive column density.
Instead, the H$^{13}$CN\,(1-0) is an ideal transition for this purpose, not only because it is unsaturated but also because its frequency is far enough away from other transitions that it is unblended.
The spectral profile of the absorption from this line is shown in panel (c) of Figure~\ref{fig:ngc3079spec}, and the systemic velocity, for which we have adopted 1147\,km\,s$^{-1}$ \citep{Koda2002}, is drawn as a dashed red line.
The solid blue line represents the continuum flux density within the 3$\arcsec$ aperture. 
The deeper absorption (the component close to the systemic velocity) is denoted ``C1''; the broader absorption component (further blue-shifted from the systemic velocity) is denoted as ``C2''. 

\subsection{Column Densities}

Assuming that local thermal equilibrium (LTE) conditions hold, the velocity integrated optical depth can be used to derive the total column density of molecules \citep{Wiklind1995}. 
For an absorption line J $\rightarrow$ J + 1:
\begin{eqnarray}
N_{total} = \frac{8\pi} {c^{3}}  \frac{\nu^{3}} {A_{J+1, J} g_{J+1}} \frac{Q(T_{ex}) exp({E_{J}/kT_{ex}})} {1 - exp({-h\nu/kT_{ex}})} \int \tau_{\nu} dV
\end{eqnarray}
where 
$A_{J+1, J}$ is the Einstein A coefficient, 
$g_{J+1}$ = 2J + 1 is the statistical weighting of level J, 
$E_{J}$ is the energy of level J, 
$Q(T_{ex})$ = $\sum_{J=0}^{\infty} g_{J} exp(-E_{J}/kT_{ex})$ is the partition function, and 
$\int \tau_{\nu} dV$ is the velocity integrated optical depth. 
In order to calculate this, we have used coefficients from the Leiden Atomic and Molecular Database \citep{Schoier2005}, for which the energy levels and radiation transitions extend up to J = 29. 
The excitation temperature, $T_{ex}$, is set as 4\,K which 
%is the lower limit to excite the transition of H$^{13}$CN\,(1-0) and 
is consistent with a high opacity of the lower transitions and, as concluded for NGC\,5128 by \cite{mul09}, is expected for densities up to $\sim10^4$\,cm$^{-3}$.
We estimate the optical depth over the source as:
\begin{eqnarray}
\tau_\nu = -ln (\frac{I_\nu^{obs}} {I_\nu^{cont}})
\end{eqnarray}
where $I_\nu^{obs}$ is the observed flux density of the absorption feature and $I_\nu^{cont}$ is the continuum flux density transmitted by the background non-thermal radio sources.

The velocity integrated optical depths of C1 and C2 are 46.13\,km\,s$^{-1}$ and 80.88\,km\,s$^{-1}$, from which we estimate associated column densities for H$^{13}$CN of $1.5 \times 10^{14}$\,cm$^{-2}$ and $2.6 \times 10^{14}$\,cm$^{-2}$ respectively.
In order to derive the hydrogen column density $N_H$, we need to apply two corrections.
The first one is for the $^{12}$C/$^{13}$C isotope ratio, which we assume to be 60, similar to the local interstellar medium \citep{Milam2005}. 
However, we note that this conversion may be underestimated if the circumnuclear region has a high star formation rate, for example M\,82 and NGC\,253 have $^{12}$C/$^{13}$C $\sim$ 100 \citep{Martin2010}.
The second correction is for the H$^{12}$CN abundance $X_{HCN}$.
\citep{Harada2010} have calculated values in the range $3\times10^{-6}$ to $9\times10^{-9}$ for gas on scales of a few to tens of parsecs around an AGN, depending on timescale.
These values are for warm ($\sim100$--400\,K) gas,  similar to the molecular gas temperatures found in the central regions of Seyfert galaxies \citep{Davies2012,hai12,Krips2008,vit14}.
In the absence of specific constraints, we adopt the geometrical mean of these two values, $X_{HCN} \sim 10^{-6.8}$, which is consistent with the plentiful evidence for X-ray enhanced HCN abundance around AGN.
Consequently, $N_H$ for the C1 and C2 components is 
$\sim 5.6 \times 10^{22}$\,cm$^{-2}$ and 
$\sim 9.8 \times 10^{22}$\,cm$^{-2}$ respectively.

Support for the high column density in the central disk of NGC\,3079 comes from the SiO absorption.
Numerous observations have shown that SiO is commonly found in massive star forming regions and in shocked clumps related to supernova remnants (e.g. \citealt{Downes1982} and \citealt{Ziurys1989a}). 
And enhancement of the SiO abundance has been predicted in high temperature chemistry reactions and models which contain shocks or molecular formation in fast molecular outflows \citep{MartinP1992,Klaassen2007}.
 
We estimate SiO column densities for components C1 and C2 in NGC\,3079 as 
$\sim 2.8 \times 10^{14}$\,cm$^{-2}$ and 
$\sim 3.4 \times 10^{14}$\,cm$^{-2}$ respectively.
For both components the ratio of SiO abundance to HCN is in the range 0.02--0.03.
These ratios are significantly higher than found in dark clouds or quiescent regions (where ratios $\la0.003$ are expected), but also significantly less than the range 0.1--1 found in massive star forming regions.
They are more comparable to the perturbed clouds associated with supernova remnant IC\,443 where $X_{SiO}/X_{HCN} \sim 0.04$--0.06 \citep{Ziurys1989b}, and also to the circumnuclear disk of NGC\,1068 where $X_{SiO}/X_{HCN} \sim 0.06$--0.1 \citep{use04}.
Adopting $X_{SiO} \sim 0.5-1\times10^{-8}$ as found by \cite{use04} for NGC\,1068, we estimate column densities of 0.3--$1\times10^{23}$\,cm$^{-2}$ for C1 and C2 in NGC\,3079.

We note that comparison with $N_H$ derived from X-ray spectra towards the AGN is difficult.
An estimate based on a BeppoSAX observation was 
$N_{H} \sim 10^{25}$\,cm$^{-2}$ \citep{Iyomoto2001}, while another based on Chandra observations was 
$N_{H} \sim 2 \times 10^{22}$\,cm$^{-2}$ \citep{Cecil2002}.
On the other hand, both X-ray observatories did strongly detect the Fe\,K$\alpha$ complex at 6.4\,keV, which is a fluorescence line produced when nuclear continuum radiation is reprocessed by circumnuclear material, and is consistent with heavy absorption.

\begin{figure*}
\begin{center}
      \includegraphics[width=140mm]{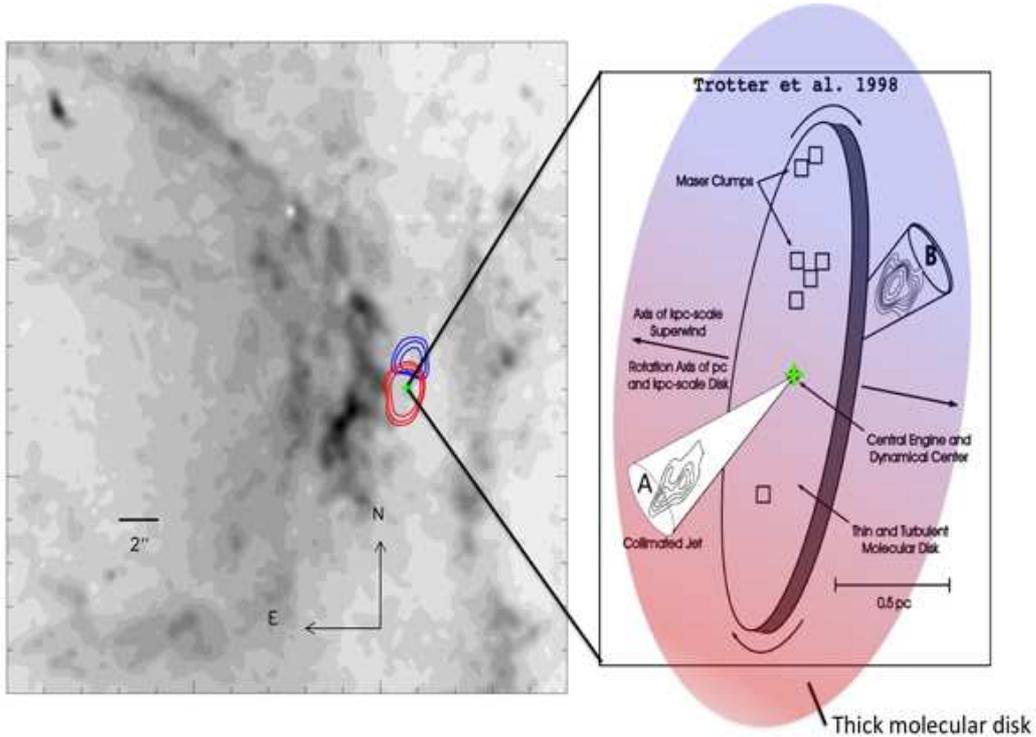}      
         \caption{Left panel: greyscale image is the [N II] + H$\alpha$ map of the supperbubble and ionized filaments observed by HST \citep{Cecil2001}. The blue and red contours are the distribution of integrated blue and red channels of H$^{12}$CN\,(1-0) presented in this paper. The central green symbol indicates the position of the nucleus. 
Right panel: the distribution of maser emitters (open squares) and 22\,GHz radio continuum (grey contours) in the nuclear region of NGC\,3079, adapted from \citet{Trotter1998} and \citet{Kondratko2005}. 
The pale blue-red colour gradient across the ellipse represents the kinematics we have observed for the rotating molecular disk. The bright radio component B in the collimated jet to the NW lies behind the disk and causes the blue-shifted absorption in the 3\,mm continuum. The cone to the SW represents the collimated jet and radio component that lies in front of the disk, and so has been left uncoloured. The green plus symbol marks the dynamical centre adopted by \citet{Kondratko2005} in their maser kinematics model.} 
\label{fig:disk_jet_filament}
\end{center} 
\end{figure*}

\subsection{Origin of the absorption}
\label{sub:origin_of_the_absorption}

The central velocities of C1 and C2 are 29\,km\,s$^{-1}$ and 137\,km\,s$^{-1}$ respectively. 
This double peaked absorption not only appears in several molecules (e.g. H$^{13}$CN, H$^{13}$CO$^+$, SiO and HN$^{13}$C) in the spectrum in Figure~\ref{fig:ngc3079spec}, but is also present in the OH absorption observed on milli-arcsec scales with the European VLBI Network (EVN) \citep{Hagiwara2004}.
Although different observations use distinct molecules, the velocity difference between the two absorption features has a consistent value of $\sim 100$\,km\,s$^{-1}$.

\subsubsection{Component C1}

Owing to the alignment of the water maser distribution with the kiloparsec-scale CO molecular structure, \citet{Trotter1998} proposed a highly inclined molecular disk, oriented roughly north-south, as the geometry of the nuclear region in NGC\,3079. 
In addition, \citet{Kondratko2005} measured the water maser kinematics and concluded the presence of a relatively thick and flared disk structure in the centre. 
We adopt these ideas from the literature and suggest that the deeper absorption C1 originates from the blue-shifted part of the rotating molecular disk, which lies in front of the bright radio jet continuum source B, which we discussed in Section~\ref{sec:Rngc3079} as also being one of the primary sources of the 3\,mm continuum. 
Adopting a black hole mass $2 \times 10^{6}$\,M$_\odot$ \citep{Kondratko2005}, and taking into account the inclination of the rotating disk and the location at which the absorption is occurring, we estimate that the velocity associated with the disk absorption should be $-47$\,km\,s$^{-1}$, in good agreement with the observed velocity of component C1.
In contrast, the red-shifted part of the rotating molecular disk is located behind the other prominent radio jet continuum source, A;
and therefore no equivalent redshifted absorption occurs in the spectrum. 
This scenario of off-centre absorption in the nuclear region of NGC\,3079 is illustrated in the right panel of Figure~\ref{fig:disk_jet_filament}. 
Based on this scheme, we argue that C1 may trace the intrinsic column density of neutral hydrogen in the edge-on rotating molecular disk, which we conclude is therefore N$_{H} \sim 6 \times 10^{22}$\,cm$^{-2}$.

Despite the size of our 3\,mm beam, this column is measured at radial scales of $\sim1$\,pc because that is the projected location of the continuum against which the absorption is occurring.
If it applies throughout the nuclear disk, we can estimate a gas mass that we can compare with the dynamical mass $M_{dyn}$ from Section~\ref{sec:model_kinematics}, and hence estimate a gas fraction.
Assuming that the absorption is caused by only 1--2 clouds along the line of sight so that no inclination correction is required, the gas mass within a radius of 112\,pc (the same as that used to derive $M_{dyn}$) is $2\times10^7$\,M$_\odot$.
For $\log{M_{dyn}} [M_\odot] = 9.16$ (Table~\ref{tab:model_int}), we find a gas fraction $f_{gas} = 1.3$\%.

These numbers are remarkably similar to the estimates of \cite{Hicks2009}.
Based on direct CO measurements as well as typical gas fractions for local spirals and star formation galaxies, these authors who argued that even with a low gas fraction of 1\% the columns exceed $10^{22}$\,cm$^{-2}$ on scales out to several tens of parsecs, and values are more typically a few $10^{23}$\,cm$^{-2}$.
Their estimates based on extinction to the stellar continuum (assuming the obscuring material is mixed with the stars) suggested slightly lower values with a mean of $2.4\times10^{22}$\,cm$^{-2}$.
Our independent measurement based on absorption confirms that in NGC\,3079, although the gas fraction in the nucleus is low, there is still sufficient material on scales of at least several parsecs to cause significant optical obscuration from some directions.

\subsubsection{Component C2}

Absorption component C2 has a broad wing that extends to at least 200\,km\,s$^{-1}$ blueward from the central velocity of C2.
Component C2 could be associated with OH absorption \citep{Hagiwara2004}.
But in contrast to the extended HCN velocity profile, the OH absorption observed with 45-mas resolution ($\sim4$\,pc) shows only the deep absorption of C2 without the broad blue-shifted wing.
We suggest below that the broad wing could be caused by absorption in outflowing material.
The H$\alpha$ features resembling outflow in the nuclear region of NGC\,3079 are displayed in the left panel of Figure~\ref{fig:disk_jet_filament}. 
The brightest emission shows that the bubble apex is close to the nucleus, and shocks induced in the bubble wall extend out to large-scales tracing the extent of the super wind \citep{Veilleux1994}. 
An ionized filament with a blue-shifted velocity of 125\,km\,s$^{-1}$ relative to systemic lies at a distance of $\sim400$\,pc (4--5\arcsec) from the nucleus, and its morphology aligns with the VLBI-scale radio jet \citep{Cecil2001}.
We speculate that the broad blue-shifted wing -- and hence the whole of component C2 -- may trace outflows on projected scales larger than 4\,pc but smaller than $\sim40$\,pc (noting that this corresponds to half the 1\arcsec\ FWHM of the IRAM beam). 
This would imply a column density in the outflowing material of 
$N_{H} \sim 10^{23}$\,cm$^{-2}$, comparable with that of the material in the disk.
While this seems remarkably high, outflowing material with a similar high $N_{H} \sim 5\times10^{22}$\,cm$^{-2}$ has been reported for the Seyfert~2 galaxy NGC\,1433 \citep{Combes2013}.

\begin{figure*}
\begin{center}
%  \hspace{-0.7cm}
%  \vspace{-0.25cm}
      \includegraphics[width=120mm]{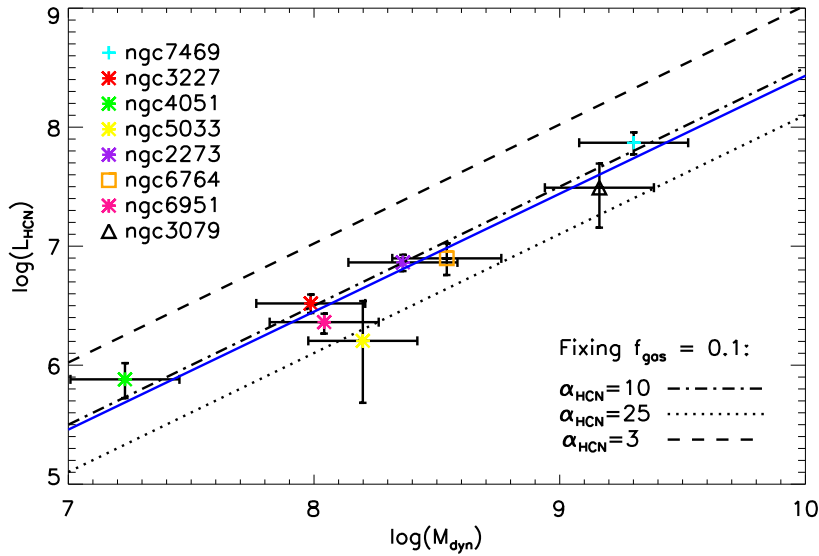}      
         \caption{HCN luminosity plotted against dynamical mass for AGN analysed here and in \citet{Sani2012}. The uncertainty in HCN luminosity includes errors from HCN flux density and distance. We adopt an uncertainty in dynamical mass of 40\% (see text in Section~\ref{sec:hcn_mdyn} for details).
The solid blue line is the best-fit linear relation for this data set.
According to Equation~\ref{eq:nvir_dynhcn}, and assuming a gas fraction of 10\%, the three black lines represent different HCN conversion factors: $\alpha_{HCN}$ = 3 (dashed), $\alpha_{HCN}$ = 10 (dot-dashed), and $\alpha_{HCN}$ = 25 (dotted)}.
\label{fig:hcndyn}
\end{center}
\end{figure*}

\section{HCN luminosity and Dynamical mass}
\label{sec:hcn_mdyn}

In this section we explore the relation between HCN luminosity $L_{HCN}$ and dynamical mass $M_{dyn}$ in the central $\sim100$\,pc of Seyfert galaxies, and its implications.
We use the combined sample of 8 Seyferts with 1\arcsec\ HCN\,(1-0) observations and kinematical modelling, as given in Table~\ref{tab:combined}.

Following \cite{Gao2007}, the HCN luminosity is calculated as 
\begin{eqnarray}
L_{HCN} =  4.1 \times 10^{3} \ S_{HCN} \Delta v \ (1+z)^{-1} \ D_{L}^{2}
\end{eqnarray}
in units of K\,km\,s$^{-1}$\,pc$^{2}$, where S$_{HCN}\Delta v$ (Jy\,km\,s$^{-1}$) is the velocity integrated HCN flux density and D$_{L}$ (Mpc) is the luminosity distance.
Since all our objects are in the very local universe, the $(1+z)^{-1}$ term is approximately unity. 
The uncertainty in HCN luminosity includes uncertainties from the velocity integrated flux density as well as the distance.

The uncertainties on $M_{dyn}$ are difficult to quantify robustly because of the assumptions and simplifications in the kinematic model from which $M_{dyn}$ is derived.
For the galaxies they modelled, \cite{Sani2012} state that the random errors are 15--20\% excluding the uncertainty for the $\sigma$ term.
From our modelling, which provides a more statistical estimate of the uncertainties based on the distribution of parameters in the best-fitting models, the uncertainties are 30--40\%.
However, we (and also \citealt{Sani2012}) adopted fixed inclinations and did not take their uncertainties into account.
The inclinations we have used for each galaxy are given in the relevant subsections of Sec.~\ref{sec:model_kinematics}, and we note that in all of the objects we have modelled as well as all of those modelled by \cite{Sani2012} the PA of the nuclear HCN emission matches rather well the PA of the host galaxy on larger scales.
Since a warp is likely to alter the PA (as well as the inclination), the similarity of the PA on large and small scales suggests the impact of any warp on the observed orientation should be modest.
We therefore have simply adopted the same inclinations as measured on large scales, noting that for NGC\,7469, \cite{Davies2004} report that a $5\degr$ change in inclination changes the dynamical mass by 17\%.
An effect of this size does not increase our uncertainties significantly, and so we adopt a global uncertainty of 40\% on all the dynamical mass estimates.

We have plotted $L_{HCN}$ against $M_{dyn}$ in Figure~\ref{fig:hcndyn}, which  shows that the relation between them is essentially linear.
The bivariate regression fit (solid blue line on the figure) is given as:
\begin{eqnarray}
\log{L_{HCN}} = (0.99\pm0.1)\log{M_{dyn}} - (1.48\pm0.84)
\label{eq:hcn_mdyn}
\end{eqnarray}
We find that Spearman's rank correlation coefficient $\rho \sim 0.90$ with 98\% significance (noting that $\rho = 1$ corresponds to two variables being monotonically related).
%The only AGN that is far off the relation is NGC\,5033, for which $L_{HCN}$ is lower than expected by a factor of $\sim3$. However, this is also associated with a large uncertainty, which is dominated by the uncertainty in the distance rather than the flux density in the line.
Since we do not know the intrinsic $L_{HCN}$ for NGC\,3079, we adopt the $S_{HCN} \Delta v$ from the absorption corrected flat-topped profile. Its $L_{HCN}$ large uncertainty is due to the difference in flux between the observed line and the absorption corrected profile.
We also looked at the surface densities for the mass and luminosity, where both axes in Figure~\ref{fig:hcndyn} are divided by the best-fit model disk area. A significant correlation still exists, indicating that disk size may not influence the relation between $L_{HCN}$ and $M_{dyn}$.
We discuss below the implication of the mass luminosity relation in terms of the gas fraction $f_{gas}$ and the conversion factor $\alpha_{HCN}$ between HCN luminosity and gas mass.

For either a single virialised (gravitationally bound) cloud, or a non-overlapping ensemble of such clouds, the ratio $\alpha_{HCN}$ between the gas mass M$_{gas}$ and the HCN luminosity L$_{HCN}$ (in units of K\,km\,s$^{-1}$\,pc$^{2}$) can be written as \citep{sol90,Downes1993,Krips2008}:
\begin{eqnarray}
\frac{M_{gas}}{L_{HCN}} \equiv \alpha_{HCN} = 2.1 \ {n_{H2}}^{0.5} \ T_{b}^{-1}
 \label{eq:vir_alpha}
\end{eqnarray}
where $n_{H2}$ (cm$^{-3}$) is the average H$_2$ number density and T$_{b}$ (K) is the brightness temperature.

However, as discussed by \cite{Downes1993} and \cite{Solomon2005}, when the line width traces the potential of a galaxy, 
i.e. $\Delta v^2 = G\,M_{dyn}/R$, 
one has to modify Equation~\ref{eq:vir_alpha} to account for the mass contributed by stars in addition to the gas.
As we have shown in Section~\ref{sec:model_kinematics}, this is the case for our observations of the HCN\,(1-0) lines; and in Section~\ref{sec:ngc6951} we show that in at least one galaxy the clouds are not self-gravitating.
In addition, we also note that since the HCN\,(1-0) transition has a critical density of $\sim3\times10^6$\,cm$^{-3}$ (e.g. \citealt{Meijerink2007}), the emission traces preferentially dense gas, whereas there may also be a significant mass of gas in less dense clumps, which might instead be traced by CO\,(1-0) which has a much lower critical density.
Following the argument given in \cite{Downes1993}, the relation between dynamical mass and HCN luminosity can be expressed as:
\begin{eqnarray}
\frac{M_{dyn}}{L_{HCN}} = 2.1 \ {n_{eq}}^{0.5} \ T_{b}^{-1}
 \label{eq:nvir_alpha}
\end{eqnarray}
where $n_{eq}$ is an equivalent H$_{2}$ number density.
Then, since $f_{gas} \equiv M_{gas}/M_{dyn} = n_{H2}/n_{eq}$, 
Equation~\ref{eq:nvir_alpha} can be rewritten as:
\begin{eqnarray}
 \frac{M_{dyn}}{L_{HCN}}  =  {f_{gas}}^{-1/2} \ \alpha_{HCN}
 \label{eq:nvir_dynhcn}
\end{eqnarray}
That there is a relation between $L_{HCN}$ and $M_{dyn}$ which is close to linear, as is apparent in Figure~\ref{fig:hcndyn}, implies either that there is an intrinsic relation between $f_{gas}$ and $\alpha_{HCN}$ or that there are typical values (to within a factor of 2 or so) for both of these quantities.

Based on a variety of sources and methods, \citet{Hicks2009} argued that the typical gas mass fraction lies in the range 4--25\%, with a typical value of $f_{gas} \sim 0.1$, within the central 200\,pc of Seyferts.
We adopt this value and, based on Equation~\ref{eq:nvir_dynhcn}, draw in Figure~\ref{fig:hcndyn} the lines corresponding to $\alpha_{HCN} = 3$ (long dash line), 10 (dot-dash line), and 25 (dotted line).
We find that $\alpha_{HCN} = 10$\,M$_{\odot}$\,(K\,km\,s$^{-1}$ pc$^{2}$)$^{-1}$ provides a remarkably good approximation to the data.
A similar HCN conversion factor $\alpha_{HCN}$ = 10$_{-3}^{+10}$ for nearby AGN (albeit with beam sizes ranging from a few arcsec up to 20\arcsec) was found via LVG analysis by \cite{Krips2008}.
As they noted, it is $\sim 2$ times smaller than the $\alpha_{HCN} = 25$ estimated by \citet{Gao2004} for nearby spiral, infrared-luminous, and ultraluminous galaxies.
However, Figure~\ref{fig:hcndyn} rules out such a high conversion factor for the centers of AGN since it would imply a gas fraction exceeding 50\%.
The difference may point towards differing excitation conditions and molecular abundances in the environments, and there is plentiful theoretical and observational evidence that X-ray excitation of gas by the AGN does have a major impact on both of these leading to an increase in the HCN luminosity \citep{Lepp1996,mal96,bog05,Meijerink2005,Meijerink2007,Kohno2003,use04,kri07,Krips2008,Davies2012}.

\section{Non self-gravitating clouds in NGC\,6951}
\label{sec:ngc6951}

\begin{figure*}
\begin{center}
  \includegraphics[width=120mm]{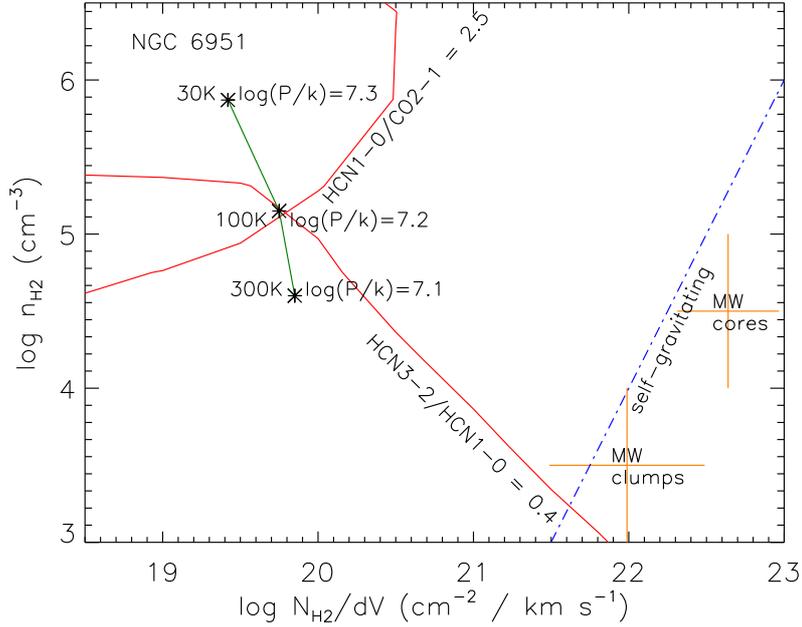}      
  \caption{Part of the 4-dimensional parameter space for LVG calculations. In this plane, the dot-dash blue line indicates the boundary for self-gravitating clouds; clumps and cores of Milky Way clouds are located to the right of this \citep{ber07}.
Red lines tracing the HCN\,(1-0)/CO\,(2-1) and HCN\,(3-2)/HCN(1-0) ratios in the centre of NGC\,6951 have been drawn for $\log{X_{HCN}/X_{CO}}=-2$ and a temperature of 100\,K (see \citealt{Davies2012}). These lines intersect at a point where the implied pressure in the cloud is $\log{P/k}\,[cm^{-3}K]=7.2$. 
Due to the well known degeneracy between density and temperature, these two parameters cannot be fully constrained. But the intersection of lines tracing the two ratios are also shown for temperatures in the range 30--300\,K (asterisks joined by solid green lines), and show that the cloud density is likely to be $\log{n}\,[cm^{-3}]=4.5$--6 
and that over this range the pressure in the clouds remains high and roughly constant.
Since all the intersections (marked by asterisks) are far to the left of the locus for self-gravitating clouds, the clouds in NGC\,6951 must be pressure confined rather than self-gravitating.}
  \label{fig:lvg6951}
\end{center}
\end{figure*}

In their high spatial resolution analyses of the Br$\gamma$ hydrogen recombination line in Seyferts, \cite{Davies2007} and \cite{hic13} were able to show that there was no on-going star formation in the central $\sim100$\,pc.
And as part of their analysis of the HCN line kinematics, \cite{Sani2012} found that while there was often star formation occurring in circumnuclear rings at radii $\ga$100\,pc, on smaller scales the evidence indicates that the star formation rates were much lower.
Our aim in this section is to assess whether the clouds in the central region are self-gravitating.
If they are not, this could be one reason why star formation appears to be suppressed on these scales.

We focus on NGC\,6951 which is one of the Seyfert galaxies in the combined sample summarised in Table~\ref{tab:combined}, and for which suitable data are available.
Observations reported by \cite{kri07} provide the HCN\,(1-0)/CO\,(2-1) ratio, which is 2.5 (if units for line fluxes are K\,km\,s$^{-1}$; it is 0.37 if the line flux units are Jy\,km\,s$^{-1}$) as discussed by \cite{Davies2012}.
These data have similar small beams and so are spatially resolved, which is important since the distributions of the two lines are very different. 
Here we use the line ratio corresponding to the central component only.
In addition, \cite{Krips2008} reported the HCN\,(3-2)/HCN\,(1-0) line ratio as 0.4 corrected for beam filling factors.
Although the beam for these data is larger (and specifically, the HCN\,(1-0) measurement used for this line ratio is a single dish measurement rather than the interferometric measurement used for the other line ratio), it is less critical since both lines are from HCN and so are expected to have more similar distributions.
Since the beam includes the circumnuclear ring, there is some uncertainty associated with whether the line ratio reflects that in the nucleus.
However, a similar measurement by \cite{Krips2008} on NGC\,1068 with a smaller beam that probes only regions well within the circumnuclear ring, yielded a HCN\,(3-2)/HCN\,(1-0) ratio of 0.21 (as such, given the HCN\,(1-0)/CO\,(2-1) ratio from \cite{use04} discussed by \citealt{Davies2012}, the analysis below could equally apply to NGC\,1068);
and a ratio of 0.15 has been reported for the nucleus of NGC\,1097 by \cite{hsi12}.
While these two galaxies could also be used for this analysis, we have not included them here.
This is partly because they are not part of our combined sample.
For NGC\,1068 it is also because the nuclear region of NGC\,1068 is very complex, having a circumnuclear bar feeding gas in to the nucleus while an outflow from the AGN disrupts the disk (\citealt{muls09,kri11,gar14}).
For NGC\,1097, it is also because the AGN is very weak and cannot really be classified as a Seyfert. 
As can be seen in Figure~\ref{fig:lvg6951} even a large uncertainty in the HCN\,(3-2)/HCN\,(1-0) ratio does not alter the general conclusion that the clouds cannot be self-gravitating.

In order to derive the physical conditions under which these line ratios can arise, we make use of the Large Velocity Gradient (LVG) approximation \citep{sob57}, which has been widely used in the literature to study gas conditions and excitiation in normal and active galaxies.
It is important to realise that LVG calculations apply to individual clouds.
On the other hand, the observations we use are of an ensemble of clouds.
However, under the usual assumption that all the clouds in the ensemble are the same, the line ratios will not depend on the number of clouds: they will be the same for a single cloud and for the ensemble.
We note, however, that this is not the case for the line width, which is very different for a single cloud and for the ensemble.
As such, to interpret the modelled ratio $N_{H_2}/dV$, or equivalently $n_{H_2}/(dV/dr)$, in the context of the observations would require additional assumptions that one sees all the clouds and that their individual line profiles and relative velocities combine to match the observed line profile.
However, our results are not affected by this issue because our comparison of the LVG calculations to observations is based solely on line ratios.

We use the LVG calculations performed by \cite{Davies2012} (we refer to that paper for details of the code and comparison to the commonly used RADEX) and extended them to include the HCN\,(3-2)/HCN\,(1-0) line ratio.
The calculations cover a wide range of parameters:
kinetic temperature $30 \leq T\,[K] \leq 300$,
HCN to CO abundance ratio $10^{-5} \leq X_{HCN}/X_{CO} \leq 10^{-2}$ 
(with $X_{CO}=10^{-4}$ as the CO abundance relative to hydrogen),
H$_2$ volume gas density $10^{3} \leq n_{H_2}\,[cm^{-3}] \leq 10^{7}$,
and a ratio of gas-density to velocity-gradient, or equivalently
column density to linewidth, of  
$5\times10^{17} \leq N_{H_2}/dV\, [cm^{-2}\,(km\,s^{-1})^{-1}] \leq 5\times10^{23}$.
One plane of this parameter space is shown in Figure~\ref{fig:lvg6951}.

For reference, we show the boundary (dot-dash blue line), to the right of which clouds are self-gravitating (see \citealt{gol01} and \citealt{Davies2012}).
For virialised clouds $\Delta V/R \sim n^{0.5}$ where $\Delta V$ is the velocity dispersion, $R$ is the cloud radius, and $n$ is the gas volume density.
The self-gravitating boundary line is found by treating $\Delta V/R$ as a velocity gradient, which gives $dV/dr \propto n_{H_2}^{0.5}$ as shown.

We have also marked the regions where one expects to find Galactic clumps and cores, using the typical properties summarised by \cite{ber07} and estimating $N_{H_2}/dV$ from their density, size, and velocity width.
Galactic clouds, with densities of $n_{H_2} = 50$--500\,cm$^{-3}$, are outside the range of the plot.
Their sizes of 2--15\,pc and velocity widths of 2--5\,km\,s$^{-1}$ indicate they have $\log{N_{H_2}/dV} \sim 20.9$ and so lie just left of the extension of the self-gravitating line.

We now discuss how, although the LVG calculation formally has 5 free parameters, we can reach a robust conclusion with only 2 line ratios.
The first parameter is the CO abundance $X_{CO}$, but because this hardly varies for a wide variety of conditions we fix it at $X_{CO} = 10^{-4}$.
The HCN abundance $X_{HCN}$ is also a parameter.
But it was already shown by \cite{Davies2012} that the  HCN\,(1-0)/CO\,(2-1) ratio alone requires an unusually high HCN abundance, $X_{HCN} \ga 10^{-6}$. 
Since this is already much higher than the typical abundance, although still consistent with what one might expect theoretically from calculations of equilibrium abundances in X-ray irradiated gas \citep{bog05}, we fix it at $X_{HCN} = 10^{-6}$.
The volume density $n_{H_2}$ and ratio of column density to linewidth $N_{H_2}/dV$ are the axes of the plot and therefore left as free parameters.
Finally, the temperature $T$ is unconstrained.
We have therefore plotted red curves tracing the locus of the line ratios above for NGC\,6951 assuming a temperature of 100\,K; and show in addition the location where the two curves intersect for temperatures of 30\,K and 300\,K (asterisks joined by the green curve), a range that covers the temperatures one might expect to find in the central regions of Seyferts \citep{Krips2008}.
This shows the well known degeneracy between temperature and density, but also indicates that, unless the temperature of the molecular gas significantly exceeds 300\,K, the gas density must be $\log{n_{H2}}\,[cm^{-3}] \ga 4.5$.

One very clear conclusion from the figure is that, independent of temperature and density, the locus of intersections for the line ratios lies far to the left of the region where one finds self-gravitating clouds.
Indeed, for the range of temperatures we have considered, the inferred pressure is roughly constant at $P/k = (1-2) \times10^7$\,cm$^{-3}$\,K suggesting that the clouds are likely to be pressure confined.
A similar conclusion was reached by \cite{zar14} for H\,{\sc II} regions and giant molecular clouds in the interacting Antennae galaxy system:
self-gravity only bound clouds above a certain mass threshold, and below this they must be bound by external pressure.
Intriguingly the pressure we find is comparable to that found by \cite{hec90} in the central few hundred parsecs of galaxies with starburst driven superwinds.
As such, the line ratios observed in the centre of NGC\,6951 are consistent with an environment dominated by supernovae, perhaps indicative of a young post-starbust.

\section{Conclusion}
\label{sec:conc}

We have presented 3\,mm interferometer data from the IRAM PdBI, which spatially resolves the HCN\,(1-0) and HCO$^+$\,(1-0) molecular lines in the central few arcsecs of three nearby Seyfert galaxies: NGC\,3079, NGC\,6764, and NGC\,5033. 
The main results of this study are as follows:
\begin{enumerate}
\item For these galaxies, and also NGC\,7469, we use a rotating disk model that takes into account beam smearing, to successfully match the observed line distribution and kinematics. The characteristics of NGC\,5033 can be fitted by a thin disk having $\sigma/v < 0.1$. In contrast NGC\,7469, NGC\,3079, and NGC\,6764 all favour a thick disk with $\sigma/v > 0.3$. Combining these results with the earlier study of \cite{Sani2012}, we find that in 7 out of 8 Seyferts, the HCN\,(1-0) line traces thick rather than thin disks in the central $\sim100$\,pc.

\item The spectrum of NGC\,3079 is dominated by numerous absorption lines, which are characterised by a double peak profile. The component closest to systemic is caused by absorption in the approaching side of the inner disk, which lies in front of a radio continuum knot in the north-west jet. Based on the depth of the H$^{13}$CN\,(1-0) feature (since H$^{12}$CN\,(1-0) is both saturated and blended with emission), we estimate the hydrogen column density in the disk to be 
$N_{H} \sim 6 \times 10^{22}$\,cm$^{-2}$. The other absorption component, with a broad blue-shifted wing extending to $-350$\,km\,s$^{-1}$, suggests the existence of a dense nuclear outflow. Correcting the HCN\,(1-0) and HCO$^+$\,(1-0) lines for continuum absorption indicates that the emission lines are also partially self-absorbed.

\item We find a relation between HCN luminosity and dynamical mass. This implies either a relation between, or typical values for, the gas fraction and the conversion factor $\alpha_{HCN}$ between HCN luminosity and gas mass. A gas fraction of $f_{gas} \sim 0.1$ and $\alpha_{HCN} \sim 10$ (consistent with the conclusion of \citealt{Krips2008} from LVG calculations) can account for the observed relation.

\item An analysis of the HCN\,(1-0)/CO\,(2-1) and HCN\,(3-2)/HCN\,(1-0) line ratios in NGC\,6951 indicates that the molecular gas is not in self-gravitating clouds. Instead, the clouds are likely to be pressure confined. The implied pressure of $P/k \sim 10^7$\,cm$^{-3}$\,K is comparable to that measured in the centers of superwinds, suggesting that the interstellar medium in the centre of NGC\,6951 is consistent with an environment dominated by supernova.
\end{enumerate}

\section*{Acknowledgements}
We thank the referee for providing a variety of important comments that have helped to improve the manuscript.
MY would like to thank Nadia Murillo for useful discussions.
RD thanks Katie Dodds-Eden for her initial work on this paper, particularly in the context of Section~\ref{sec:cd3079}. 
Based on observations carried out under project number U08A with the IRAM PdBI. IRAM is supported by INSU/CNRS (France), MPG (Germany) and IGN (Spain).
This research has made use of the NASA/IPAC Extragalactic Database (NED) which is operated by the Jet Propulsion Laboratory, California Institute of Technology, under contract with the National Aeronautics and Space Administration.

% Don't change these lines
\bsp	% typesetting comment
\label{lastpage}


\begin{thebibliography}{99}

\bibitem[Aalto et al.(2015)]{aal15}
Aalto S., Mart\'in S., Costagliola F., Gonz\'alez-Alfonso E., Muller S., et al., 2015,
A\&A, 584, A42

\bibitem[Antonucci(1993)]{Antonucci1993}
Antonucci, R., 1993, ARAA, 31, 473 

\bibitem[Bergin \& Tafalla(2007)]{ber07}
Bergin E., Tafalla M., 2007,
ARA\&A, 45, 339

\bibitem[Boger \& Sternberg(2005)]{bog05}
Boger G., Sternberg A., 2005, 
ApJ, 632, 302

\bibitem[Braine et al.(1997)]{bra97}
Braine J., Gu\'elin M., Dumke M., Brouillet N., Herpin F., Wielebinski R., 1997,
A\&A, 326, 963

\bibitem[Burtscher et al.(2015)]{bur15}
Burtscher L., Orban de Xivry G., Davies R., Janssen A., Lutz D., et al., 2015,
A\&A, 578, 47

\bibitem[Cecil et al.(2001)]{Cecil2001} 
Cecil, G., Bland-Hawthorn, J., Veilleux, S., \& Filippenko, A.~V., 2001, 
ApJ, 555, 338

\bibitem[Cecil et al.(2002)]{Cecil2002} 
Cecil, G., Bland-Hawthorn, J., \& Veilleux, S., 2002, 
ApJ, 576, 745

\bibitem[Cid Fernandes et al.(2004)]{cid04}
Cid Fernandes R., Gonz{\'a}lez Delgardo R., Schmitt H., Storchi-Bergmann T., Martins L., et al., 2004,
ApJ, 605, 105

\bibitem[Combes et al.(2013)]{Combes2013} 
Combes, F., Garc{\'{\i}}a-Burillo, S., Casasola, V., et al., 2013, 
A\&A, 558, A124 

\bibitem[Costagliola et al.(2011)]{Costagliola2011} 
Costagliola, F., Aalto, S., Rodriguez, M.~I., et al., 2011, 
A\&A, 528, A30 

\bibitem[Davies et al.(2004)]{Davies2004} Davies, R.~I., Tacconi, 
L.~J., \& Genzel, R., 2004, ApJ, 602, 148 

\bibitem[Davies et al.(2006)]{Davies2006} Davies, R.~I., Thomas, 
J., Genzel, R., et al., 2006, ApJ, 646, 754

\bibitem[Davies et al.(2007)]{Davies2007}
Davies, R.~I., M{\"u}ller S{\'a}nchez, F., Genzel, R., et al. 2007, 
ApJ, 671, 1388

\bibitem[Davies et al.(2011)]{Davies2011} Davies, R., F{\"o}rster 
Schreiber, N.~M., Cresci, G., et al., 2011, ApJ, 741, 69 

\bibitem[Davies et al.(2012)]{Davies2012} 
Davies, R., Mark, D., \& Sternberg, A., 2012, A\&A, 537, A133 

\bibitem[Downes et al.(1982)]{Downes1982} Downes, D., Genzel, R., 
Hjalmarson, A., Nyman, L.~A., \& Ronnang, B., 1982, ApJL, 252, L29

\bibitem[Downes et al.(1993)]{Downes1993} Downes, D., Solomon, 
P.~M., \& Radford, S.~J.~E., 1993, ApJL, 414, L13

\bibitem[Elitzur \& Shlosman(2006)]{Elitzur2006}
Elitzur, M., \& Shlosman, I., 2006, ApJL, 648, L101 

\bibitem[Engel et al.(2011)]{eng11}
Engel H., Davies R., Genzel R., Taccconi L., Sturm E., Downes D., 2011,
ApJ, 729, 58

\bibitem[Esquej et al.(2014)]{esq14}
Esquej P., Alonso-Herrero A., Gonz\'alez-Mart\'in O., H\"onig S., Hern\'an-Caballero A., et al., 2014,
ApJ, 780, 86

\bibitem[Feltre et al.(2012)]{Feltre2012} Feltre, A., 
Hatziminaoglou, E., Fritz, J., \& Franceschini, A., 2012, MNRAS, 426, 120

\bibitem[Fritz et al.(2006)]{fri06}
Fritz J., Franceschini A., Hatziminaoglou E., 2006,
MNRAS, 366, 767

\bibitem[Gao et al.(2007)]{Gao2007} Gao, Y., Carilli, C.~L., 
Solomon, P.~M., \& Vanden Bout, P.~A., 2007, ApJL, 660, L93 

\bibitem[Gao \& Solomon(2004)]{Gao2004} 
Gao, Y., \& Solomon, P.~M., 2004, ApJS, 152, 63

\bibitem[Garci\'a-Burillo et al.(2014)]{gar14}
Garci\'a-Burillo S., Combes F., Usero A., Aalto S., Krips M., et al., 2014,
A\&A, 567, A125

\bibitem[Goldsmith(2001)]{gol01}
Goldsmith P., 2001,
ApJ, 557, 736

\bibitem[Gottlieb et al.(1983)]{got83}
Gottlieb C., Gottlieb E., Thaddeus P., 1983,
ApJ, 264, 740

\bibitem[Granato \& Danese(1994)]{Granato1994} 
Granato, G.~L., \& Danese, L., 1994, MNRAS, 268, 235

\bibitem[Granato et al.(1997)]{Granato1997} 
Granato, G.~L., Danese, L., \& Franceschini, A., 1997, ApJ, 486, 147

\bibitem[Hagiwara et al.(2004)]{Hagiwara2004} 
Hagiwara, Y., Kl{\"o}ckner, H.-R., \& Baan, W., 2004, MNRAS, 353, 1055 

\bibitem[Hailey-Dunsheath et al.(2012)]{hai12}
Hailey-Dunsheath S., Sturm E., Fischer J., Sternberg A., Graci\'a-Carpio J., et al., 2012,
ApJ, 755, 57

\bibitem[Harada et al.(2010)]{Harada2010} 
Harada, N., Herbst, E., \& Wakelam, V., 2010, ApJ, 721, 1570

\bibitem[Hawarden et al.(1995)]{haw95}
Hawarden T., Israel F., Geballe T., Wade R., 1995,
MNRAS, 276, 1197

\bibitem[Heckman et al.(1990)]{hec90}
Heckman T., Armus L., Miley G., 1990,
ApJSS, 74, 833

\bibitem[Hicks et al.(2009)]{Hicks2009} 
Hicks, E.~K.~S., Davies, R.~I., Malkan, M.~A., et al., 2009, ApJ, 696, 448

\bibitem[Hicks et al.(2013)]{hic13} 
Hicks, E.~K.~S., Davies, R.~I., Maciejewski W., Emsellem E., Malkan M., et al. 2013,
ApJ, 768, 107

\bibitem[H\"onig et al.(2006)]{hoe06}
Hoenig S., Beckert T., Ohnaka K., Weigelt G., 2006,
A\&A, 452,459

\bibitem[Hopkins et al.(2012)]{Hopkins2012} Hopkins, P.~F., 
Hayward, C.~C., Narayanan, D., \& Hernquist, L., 2012, MNRAS, 420, 320

\bibitem[Hota \& Saikia(2006)]{hot06}
Hota A., Saikia D., 2006,
MNRAS, 371, 945

\bibitem[Hsieh et al.(2012)]{hsi12}
Hsieh P.-Y., Ho P., Kohno K., Hwang C.-Y., Matsushita S., 2012,
ApJ, 747, 90

\bibitem[Huchra et al.(1995)]{huc95}
Huchra J., Geller M., Corwin H., 1995
ApJSS, 99, 391

\bibitem[Iyomoto et al.(2001)]{Iyomoto2001} Iyomoto, N., Fukazawa, 
Y., Nakai, N., \& Ishihara, Y., 2001, ApJL, 561, L69 

\bibitem[Kharb et al.(2010)]{Kharb2010} Kharb, P., Hota, A., 
Croston, J.~H., et al., 2010, ApJ, 723, 580

\bibitem[Klaas \& Walker(2002)]{kla02}
Klaas U., Walker H., 2002,
A\&A, 391, 911

\bibitem[Klaassen \& Wilson(2007)]{Klaassen2007} 
Klaassen, P.~D., \& Wilson, C.~D., 2007, ApJ, 663, 1092

\bibitem[Koda et al.(2002)]{Koda2002}
Koda, J., Sofue, Y., Kohno, K., et al., 2002, 
ApJ, 573, 105

\bibitem[Kohno et al.(2003)]{Kohno2003} 
Kohno, K., Vila-Vilar{\'o}, B., Sakamoto, S., et al., 2003, PASJ, 55, 103 

\bibitem[Kohno(2005)]{koh05}
Kohno K., 2005,
in {\em The Evolution of Starbursts: The 331st Wilhelm and Else Heraeus Seminar},
AIP Conf. Proc. vol. 783, pp. 203--208

\bibitem[Kohno et al.(2008)]{Kohno2008}
Kohno, K., Nakanishi, K., Tosaki, T., et al., 2008, Ap\&SS, 313, 279

\bibitem[Kondratko et al.(2005)]{Kondratko2005} Kondratko, P.~T., 
Greenhill, L.~J., \& Moran, J.~M., 2005, ApJ, 618, 618 

\bibitem[Krips et al.(2007)]{kri07}
Krips, M., Neri, R., Garc\'a-Burillo, S., et al., 2007, 
A\&A, 468, L63

\bibitem[Krips et al.(2008)]{Krips2008}
Krips, M., Neri, R., Garc{\'{\i}}a-Burillo, S., et al., 2008, 
ApJ, 677, 262 

\bibitem[Krips et al.(2011)]{kri11}
Krips, M., Mart\'in S., Eckart A., Neri R., Garci\'a-Burillo S., et al., 2011,
ApJ, 736, 37

\bibitem[Krolik \& Begelman(1988)]{Krolik1988} 
Krolik, J.~H., \& Begelman, M.~C., 1988, 
ApJ, 329, 702

\bibitem[Kukula et al.(1995)]{kuk95}
Kukula M., Pedlar A., Baum S., O'Dea C., 1995,
MNRAS, 276, 1262

\bibitem[Leon et al.(2007)]{Leon2007} 
Leon, S., Eckart, A., Laine, S., et al., 2007, 
A\&A, 473, 747 

\bibitem[Lepp \& Dalgarno(1996)]{Lepp1996} Lepp, S., \& Dalgarno, A., 1996, 
A\&A, 306, L21

\bibitem[Maloney et al.(1996)]{mal96}
Maloney P., Hollenbach D., Tielens A., 1996,
ApJ, 466, 561

\bibitem[Mart{\'{\i}}n et al.(2010)]{Martin2010} 
Mart{\'{\i}}n, S., Aladro, R., Mart{\'{\i}}n-Pintado, J., \& Mauersberger, R., 2010, 
A\&A, 522, A62 

\bibitem[Martin-Pintado et al.(1992)]{MartinP1992} 
Martin-Pintado, J., Bachiller, R., \& Fuente, A., 1992, 
A\&A, 254, 315

\bibitem[Meijerink \& Spaans(2005)]{Meijerink2005} 
Meijerink, R., \& Spaans, M., 2005, 
A\&A, 436, 397 

\bibitem[Meijerink et al.(2007)]{Meijerink2007} 
Meijerink, R., Spaans, M., \& Israel, F.~P., 2007, 
A\&A, 461, 793 
%\bibitem[Middelberg et al.(2005)]{Middelberg2005} Middelberg, E., 
%Krichbaum, T.~P., Roy, A.~L., Witzel, A., 
%\& Zensus, J.~A.\ 2005, Future Directions in High Resolution Astronomy, 340, 140 

\bibitem[Middelberg et al.(2007)]{Middelberg2007}
Middelberg E., Agudo I., Roy A., Krichbaum T., 2007,
MNRAS, 377, 731

\bibitem[Milam et al.(2005)]{Milam2005} 
Milam, S.~N., Savage, C., Brewster, M.~A., Ziurys, L.~M., \& Wyckoff, S., 2005, 
ApJ, 634, 1126

\bibitem[Morganti et al.(2007)]{mor07}
Morganti R., Holt J., Saripalli L., Oosterloo T., Tadhunter Cl., 2007,
A\&A, 476, 735

\bibitem[Morganti(2012)]{mor12}
Morganti R., 2012
in {\em Nuclei of Seyfert galaxies and QSOs -- Central engine \& conditions of star formation}, Proceedings of Science, pub. Max-Planck-Insitut f\"ur Radioastronomie (MPIfR), Bonn, Germany; online at  http://pos.sissa.it/cgi-bin/reader/conf.cgi?confid=169

\bibitem[Morganti et al.(2015)]{mor15}
Morganti R., Oosterloo T., Oonk J., Frieswijk W., Tadhunter Cl., 2015,
A\&A, 580, 1

\bibitem[Muller \& Dinh-V-Trung(2009)]{mul09}
Muller S., Dinh-V-Trung, 2009,
ApJ, 696, 176

\bibitem[M{\"u}ller-S{\'a}nchez et al.(2009)]{muls09} 
M{\"u}ller-S{\'a}nchez, F., Davies R., Genzel R., Tacconi L., Eisenhauer F., Hicks E., Friedrich S., Sternberg A., 2009,
ApJ, 691, 749

\bibitem[M{\"u}ller-S{\'a}nchez et al.(2013)]{Muller2013} 
M{\"u}ller-S{\'a}nchez, F., Prieto, M.~A., Mezcua, M., et al., 2013, 
ApJL, 763, L1 

\bibitem[Nenkova et al.(2002)]{Nenkova2002} 
Nenkova, M., Ivezi{\'c}, {\v Z}., \& Elitzur, M., 2002, 
ApJL, 570, L9

\bibitem[Netzer(2015)]{net15}
Netzer H., 2015,
ARA\&A, 53, 365

\bibitem[Papadopoulos(2007)]{Papadopoulos2007} 
Papadopoulos, P.~P., 
2007, ApJ, 656, 792 

\bibitem[P{\'e}rez-Torres \& Alberdi(2007)]{Perez2007} 
P{\'e}rez-Torres, M.~A., \& Alberdi, A., 2007, 
MNRAS, 379, 275 

\bibitem[Pier \& Krolik(1992)]{pie92}
Pier E., Krolik J., 1992,
ApJ, 401, 99

\bibitem[Rangwala et al.(2015)]{ran15}
Rangwala N., Maloney P., Wilson C., Glenn J., Kamenetzky J., Spinoglio L., 2015,
ApJ, 806, 17

\bibitem[Riffel et al.(2009)]{rif09}
Riffel R., Postoriza M., Rodr\'iguez-Ardila A., Bonatto C., 2009
MNRAS, 400, 273
%\bibitem[Rodr{\'{\i}}guez-Ardila 
%\& Mazzalay(2006)]{RM2006} Rodr{\'{\i}}guez-Ardila, A., \& Mazzalay, X.\ 2006, MNRAS, 367, L57 

\bibitem[Sakamoto et al.(2009)]{sak09}
Sakamoto K., Aalto S., Wilner D., Black J., Conway J., et al., 2009,
ApJ, 700, L104

\bibitem[Sani et al.(2012)]{Sani2012} 
Sani, E., Davies, R.~I., Sternberg, A., et al. 2012, 
MNRAS, 424, 1963 

\bibitem[Schartmann et al.(2005)]{sch05}
Schartmann M., Meisenheimer K., Camenzind M., Wolf S., Henning T., 2005,
A\&A, 437, 861

\bibitem[Schartmann et al.(2008)]{sch08}
Schartmann M., Meisenheimer K., Camenzind M., Wolf S., Tristram K., Henning T., 2008,
A\&A, 482, 67

\bibitem[Schartmann et al.(2014)]{sch14}
Schartmann M., Wada K., Prieto M.A., Burkert A., Tristram K., 2014,
MNRAS, 445, 3878

\bibitem[Schweitzer et al.(2008)]{Schweitzer2008} 
Schweitzer, M., Groves, B., Netzer, H., et al., 2008, ApJ, 679, 101

\bibitem[Sch{\"o}ier et al.(2005)]{Schoier2005}
Sch{\"o}ier, F.~L., van der Tak, F.~F.~S., van Dishoeck, E.~F., \& Black, J.~H., 2005, 
A\&A, 432, 369 

\bibitem[Sobolev(1957)]{sob57}
Sobolev V., 1957,
SvA, 1, 678

\bibitem[Sofue \& Rubin(2001)]{Sofue2001} 
Sofue, Y., \& Rubin, V., 2001, ARAA, 39, 137 

\bibitem[Solomon et al.(1990)]{sol90}
Solomon P., Radford S., Downes D., 1990,
ApJL, 348, 53

\bibitem[Solomon \& Vanden Bout (2005)]{Solomon2005} 
Solomon P.M. \& Vanden Bout P.A., 2005, 
ARAA, 43, 677

\bibitem[Sternberg \& Dalgarno(1995)]{Sternberg1995} 
Sternberg, A., \& Dalgarno, A., 1995, 
ApJS, 99, 565

\bibitem[Stevens \& Gear(2000)]{ste00}
Stevens J., Gear W., 2000,
MNRAS, 312, L5
%\bibitem[Tacconi et al.(1994)]{Tacconi1994} Tacconi, L.~J., Genzel, 
%R., Blietz, M., et al.\ 1994, ApJL, 426, L77 

\bibitem[Takamiya \& Sofue(2002)]{Takamiya2002} 
Takamiya, T., \& Sofue, Y., 2002, 
ApJL, 576, L15 

\bibitem[Thean et al.(1997)]{Thean1997}
Thean A., Mundell C., Pdelar A., \& Nicholson R., 1997,
MNRAS, 290, 15

\bibitem[Thompson et al.(2005)]{Thompson2005} 
Thompson, T.~A., Quataert, E., \& Murray, N., 2005, 
ApJ, 630, 167 

\bibitem[Tristram et al.(2014)]{Tristram2014} 
Tristram, K.~R.~W., Burtscher, L., Jaffe, W., et al., 2014, 
A\&A, 563, AA82

\bibitem[Trotter et al.(1998)]{Trotter1998} 
Trotter, A.~S., Greenhill, L.~J., Moran, J.~M., et al., 1998, 
ApJ, 495, 740 

\bibitem[Urry \& Padovani(1995)]{UP1995} 
Urry, C.~M., \& Padovani, P., 1995, 
PASP, 107, 803

\bibitem[Usero et al.(2004)]{use04}
Usero A., Garc\'ia-Burillo S., Fuente A., Mart\'in-Pintado J., Rodr\'iguez-Fern\'andez N., 2004,
A\&A, 419, 897

\bibitem[Veilleux et al.(1994)]{Veilleux1994} 
Veilleux, S., Cecil, G., Bland-Hawthorn, J., et al., 1994, 
ApJ, 433, 48 

\bibitem[Viti et al.(2014)]{vit14}
Viti S., Garc\'ia-Burillo S., Fuente A.. Hunt L., Usero A., et al. 2014,
A\&A, 570, A28

\bibitem[Vollmer et al.(2008)]{vol08}
Vollmer B., Beckert T., Davies R., 2008,
A\&A, 491, 441

\bibitem[Wada \& Norman(2002)]{Wada2002} 
Wada, K., \& Norman, C.~A., 2002, 
ApJL, 566, L21

\bibitem[Wada et al.(2009)]{wad09}
Wada K., Papadopoulos P., Spaan M., 2009,
ApJ, 702, 63

\bibitem[Wada(2012)]{wad12}
Wada K., 2012,
ApJ, 758, 66

\bibitem[White et al.(1997)]{whi97}
White R., Becker R., Helfand D., Gregg M., 1997,
ApJ, 475, 479

\bibitem[Wiklind \& Combes(1995)]{Wiklind1995} 
Wiklind, T., \& Combes, F., 1995, 
A\&A, 299, 382 

\bibitem[Yamagishi et al.(2010)]{yam10}
Yamagishi M., Kandea H., Ishihara D., Komugi S., Suzuki T., Onaka T., 2010,
PASJ, 62, 1085

\bibitem[Zaragoza-Cardiel et al.(2014)]{zar14}
Zaragoza-Cardiel J., Font J., Beckman J., Garc\'ia-Lorenzo B., Erroz-Ferrer S., Guti\'errez L., 2014
MNRAS, 445, 1412

\bibitem[Ziurys et al.(1989a)]{Ziurys1989a} 
Ziurys, L.~M., Snell, R.~L., \& Dickman, R.~L., 1989, 
ApJ, 341, 857 

\bibitem[Ziurys et al.(1989b)]{Ziurys1989b} 
Ziurys, L.~M., Friberg, P., \& Irvine, W.~M., 1989, 
ApJ, 343, 201

\end{thebibliography}
\end{document}